\documentclass{article}
\usepackage{arxiv}

\usepackage{subfig}
\usepackage{rotating}
\usepackage{placeins}
\usepackage{stfloats}
\usepackage{trsym}
\usepackage{amsmath}
\usepackage{cite}
\usepackage{enumitem}
\usepackage{url}
\usepackage{xpatch}
\usepackage{multirow}
\usepackage{svg}
\usepackage{tabularx}
\usepackage{algorithm2e}
\usepackage{footnote}
\usepackage{booktabs}
\makesavenoteenv{tabular}
\makesavenoteenv{table}

\usepackage{listings}
\usepackage{color}

\usepackage{xcolor}

\colorlet{punct}{red!60!black}
\definecolor{background}{HTML}{EEEEEE}
\definecolor{delim}{RGB}{20,105,176}
\colorlet{numb}{magenta!60!black}

\definecolor{dkgreen}{rgb}{0,0.6,0}
\definecolor{gray}{rgb}{0.5,0.5,0.5}
\definecolor{mauve}{rgb}{0.50,0,0.80}
\lstdefinelanguage{json}{
    commentstyle=\color{eclipseStrings}, 
    stringstyle=\color{mauve}, 
    basicstyle=\small\ttfamily,
    numbers=left,
    numberstyle=\small,
    stepnumber=1,
    numbersep=8pt,
    showstringspaces=false,
    breaklines=true,
    frame=lines,
    string=[s]{"}{"},
    literate=
     *{:}{{{\color{dkgreen}{:}}}}{1}
      {,}{{{\color{dkgreen}{,}}}}{1}
      {\{}{{{\color{delim}{\{}}}}{1}
      {\}}{{{\color{delim}{\}}}}}{1}
      {[}{{{\color{delim}{[}}}}{1}
      {]}{{{\color{delim}{]}}}}{1},
}

\begin{document}
\renewcommand{\baselinestretch}{1.20}
\title{FOS: A Modular FPGA Operating System for Dynamic Workloads}

%
\author{Anuj Vaishnav,\\
    Department of Computer Science,\\
    The University of Manchester, \\
    \texttt{{anuj.vaishnav}@manchester.ac.uk}
    \And
    Khoa Dang Pham,\\
    Department of Computer Science,\\
    The University of Manchester, \\
    \texttt{{khoa.pham}@manchester.ac.uk}
    \And
    Joseph Powell,\\
    Department of Computer Science,\\
    The University of Manchester, \\
    \texttt{{joseph.powell}@manchester.ac.uk}
    \And
    Dirk Koch,\\
    Department of Computer Science,\\
    The University of Manchester, \\
    \texttt{{dirk.koch}@manchester.ac.uk}
}
\maketitle

\begin{abstract}

With FPGAs now being deployed in the cloud and at the edge, there is a need for scalable design methods which can incorporate the heterogeneity present in the hardware and software components of FPGA systems. 
Moreover, these FPGA systems need to be maintainable and adaptable to changing workloads while improving accessibility for the application developers. 
However, current FPGA systems fail to achieve modularity and support for multi-tenancy due to dependencies between system components and lack of standardised abstraction layers. 
To solve this, we introduce a modular FPGA operating system -- FOS, which adopts a modular FPGA development flow to allow each system component to be changed and be agnostic to the heterogeneity of EDA tool versions, hardware and software layers. 
Further, to dynamically maximise the utilisation transparently from the users, FOS employs resource-elastic scheduling to arbitrate the FPGA resources in both time and spatial domain for any type of accelerators.
Our evaluation on different FPGA boards shows that FOS can provide performance improvements in both single-tenant and multi-tenant environments while substantially reducing the development time and, at the same time, improving flexibility.
\end{abstract}


\maketitle

\section{Introduction}
\label{sec:introduction}

FPGA accelerators can provide high-performance computing at very low energy cost for applications ranging from neural-networks to network processing. This has brought FPGAs in cloud datacenters as well as in embedded systems at the edge. 
However, to sustain this large scope of requirements present in terms of heterogeneity of devices, environments, EDA tools, users and developer needs, we require a standardised way to manage and integrate FPGA systems components, in other words: we need an FPGA operating system.


The idea of an FPGA operating system is old and has taken many forms over time. It has evolved from providing a)~OS-level APIs such as hardware threads~\cite{reconOS, hthreads} and UNIX interfaces~\cite{borph} for allowing hardware accelerators to use existing software OS services, to b)~light-weight FPGA shells with library APIs for hardware acceleration~\cite{pvfpga, networkAttached, byma, khawaja2018armophos, Chen:2014}. 

\begin{figure}
    \centering
    \includegraphics[width=0.86\linewidth]{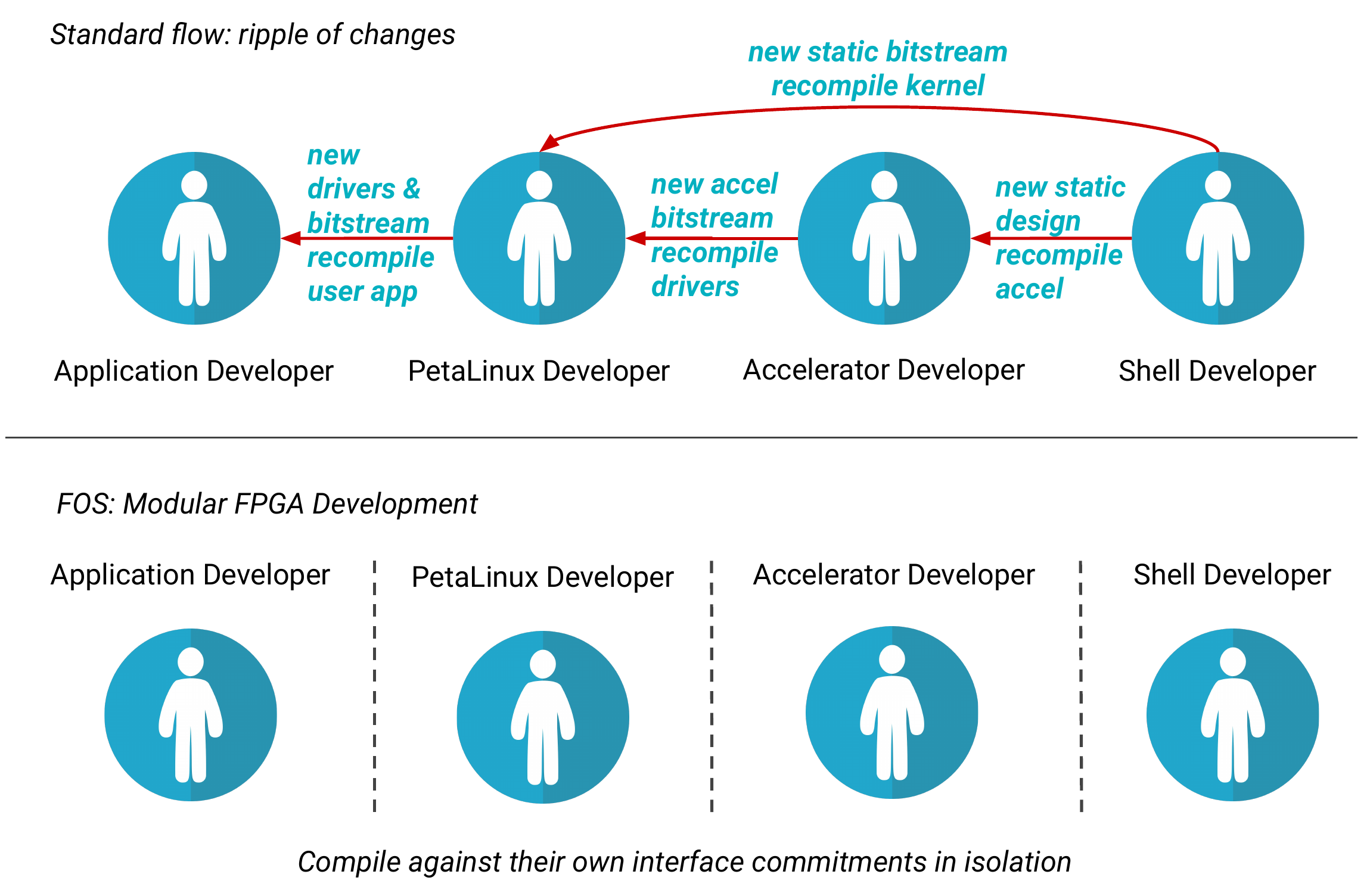}
    \caption{If we update the shell IP or EDA tool version, the rest of the system components must be recompiled because of the dependencies in standard tool flow (see Section~\ref{sec:modular_dev_flow}). Ideally, we want each component to be compiled in isolation with given interfaces, as we do for FOS.}
    \label{fig:modular_dev_example}
\end{figure}

We believe this evolution continues for two primary reasons: \textit{overhead and lack of portability}. 
The introduction of intermediate layers and communication cost must be low in terms of latency as well as in the amount of resources required. 
However, to achieve this is difficult because hardware accelerators and boards require unique optimisations and implementations, leading to heterogeneity in FPGA shells, accelerators, EDA tools, and the low-level software required to interface with it. In particular, this reduces not only the portability of a system but also its \textit{maintainability}, which is a key aspect for any operating system (OS). When using, for example, the current tool flows of the major FPGA vendors, a simple change in tool version or addition of system IP can lead to re-compilation of the whole FPGA stack~\cite{sdaccel, sdsoc}, as shown in Figure~\ref{fig:modular_dev_example}. Essentially, implying that an operating system update means recompilation of all applications which run on top of it.

Moreover, with the need for multi-tenancy in dynamic environments like the cloud, we expect multiple types of workloads and accelerators to execute concurrently~\cite{virtualization_survey} (e.g., for enabling resource pooling on FPGA accelerator cards). However, most present FPGA systems operate and support either static accelerators or single accelerator execution~\cite{aws, catapult, xilinx2017clouds, Fahmy}. Systems which support hosting multiple accelerators, do not allow resources to be re-allocated during execution or employ only time-domain multiplexing~\cite{vesper2017pciehls, xia_kerone, Happe2015MultitaskingReconOS}. This commonly leads to under-utilisation and a lack of \textit{adaptability} in the system. 


At the same time, there is a need for better APIs to make FPGAs more \textit{accessible} to software programmers or application domain experts that are commonly not FPGA experts. This should be achieved, while ensuring that hardware developers can integrate their accelerators into software stacks with high productivity, i.e. without changing their hardware designs or writing complex software wrappers. 

\begin{figure}
    \centering
    \includegraphics[width=0.75\linewidth]{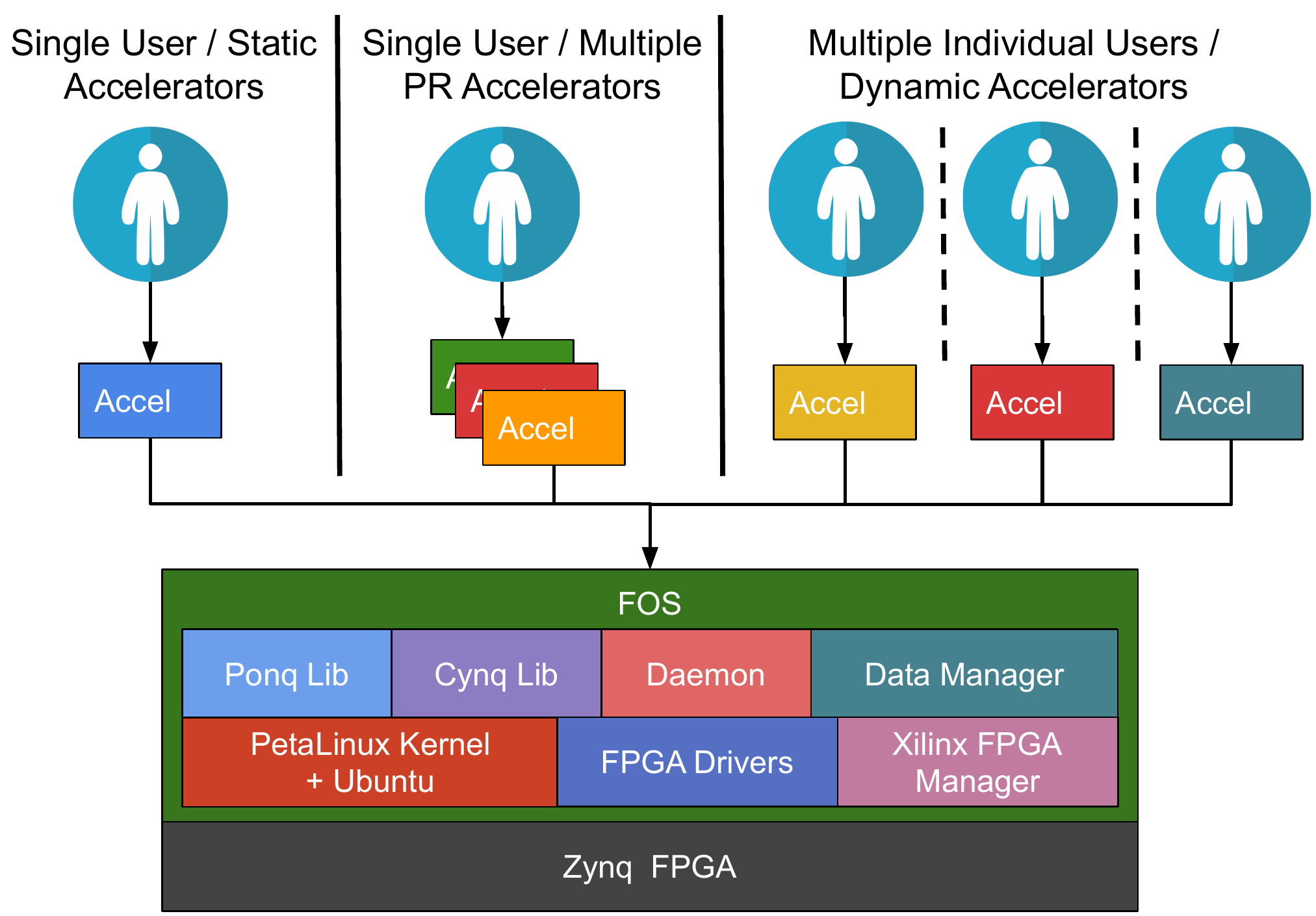}
    \caption{System components and usage mode of the FPGA Operating System -- FOS.}
    \label{fig:fos_usage_mode}
\end{figure}

To solve these challenges of maintainability, adaptability, and accessibility, we are introducing a modular FPGA Operating System (FOS). FOS adopts a modular FPGA development flow to allow each type of OS component (hardware or software) to be replaced, reused or ported to different FPGA systems without extensive effort or compilation time. Moreover, the runtime system provides full support for multi-tenancy with the ability to concurrently execute accelerators written in various languages (C, C++, OpenCL, or RTL) and dynamically replicate or switch to a different version of an accelerator to improve utilisation and system performance.
To interact with these accelerators, application developers can use high-level APIs (available in multiple languages) to access the FPGA in three modes: 1)~static acceleration for a single user, 2)~dynamic acceleration for a single user and 3)~dynamic acceleration in multi-tenant environments, as shown in Figure~\ref{fig:fos_usage_mode}. At the same time, hardware developers can write light-weight interface descriptions to integrate their hardware accelerators into the FOS platform and benefit from its high-level software APIs.


All the research artefacts proposed in this paper were presented at the FPL 2019 Demo Night and are available online under an open-source license\footnote{https://github.com/khoapham/fos}. 
The key contributions of this paper are:
\begin{itemize}
    \item A \textit{modular FPGA development flow} to encapsulate the heterogeneity at each level of the development process (Section~\ref{sec:modular_dev_flow}).
    \item An \textit{open-source software platform} to improve programmability for static acceleration systems, dynamic systems for single-user as well as multi-user dynamic acceleration systems (Section~\ref{sec:fos_overview} and~\ref{sec:implementation}). 
    \item A \textit{decoupled compilation flow for shell and modules} with support for module relocation and flexibility to combine PR regions to host accelerators of varying sizes (Section~\ref{sec:decoupled_compilation_flow}).
    \item A \textit{resource-elastic runtime system for dynamic workloads} supporting virtually any type of accelerators (written in RTL, C, C++, and OpenCL) with a unified user interface and programming model (Section~\ref{sec:fos_runtime}). 
    \item A thorough evaluation of the FPGA Operating System (FOS) on resource overhead, compilation latency, memory throughput performance, application performance, as well as flexibility and maintainability of the system (Section~\ref{sec:evaluation}). 
\end{itemize}

\section{Concepts}
\subsection{FPGA Operating System}
\label{sec:fpga_os}

\begin{figure}
    \centering
    \includegraphics[width=0.85\linewidth]{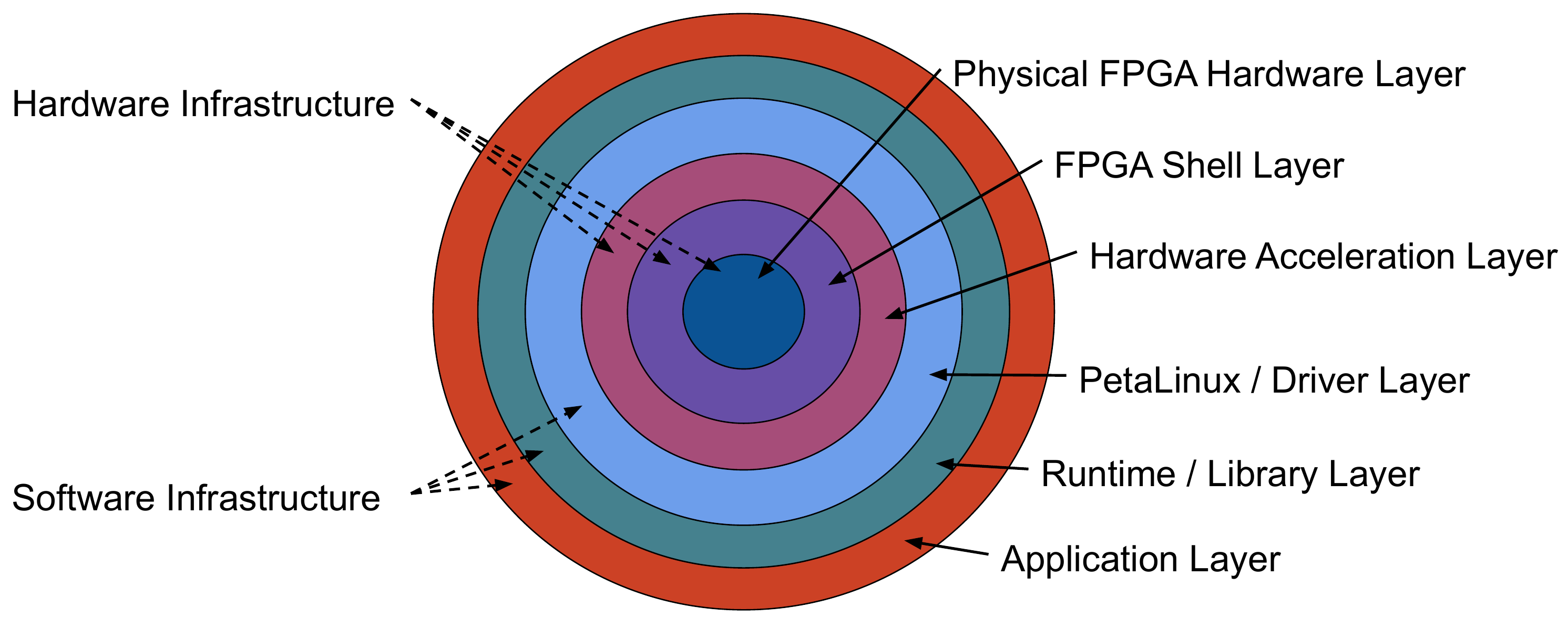}
    \caption{The layered architecture of an FPGA operating system.}
    \label{fig:fpga_os_layers}
\end{figure}

Unlike standard CPU operating systems which consist only of a software infrastructure, an FPGA operating system requires an infrastructure at both the hardware and the software level (as shown in Figure~\ref{fig:fpga_os_layers}). The following subsections describe the role of each level in detail.

\subsubsection{Hardware Infrastructure}

\begin{figure}
	\centering
	\includegraphics[width=.75\columnwidth,trim=10 10 10 10,clip]{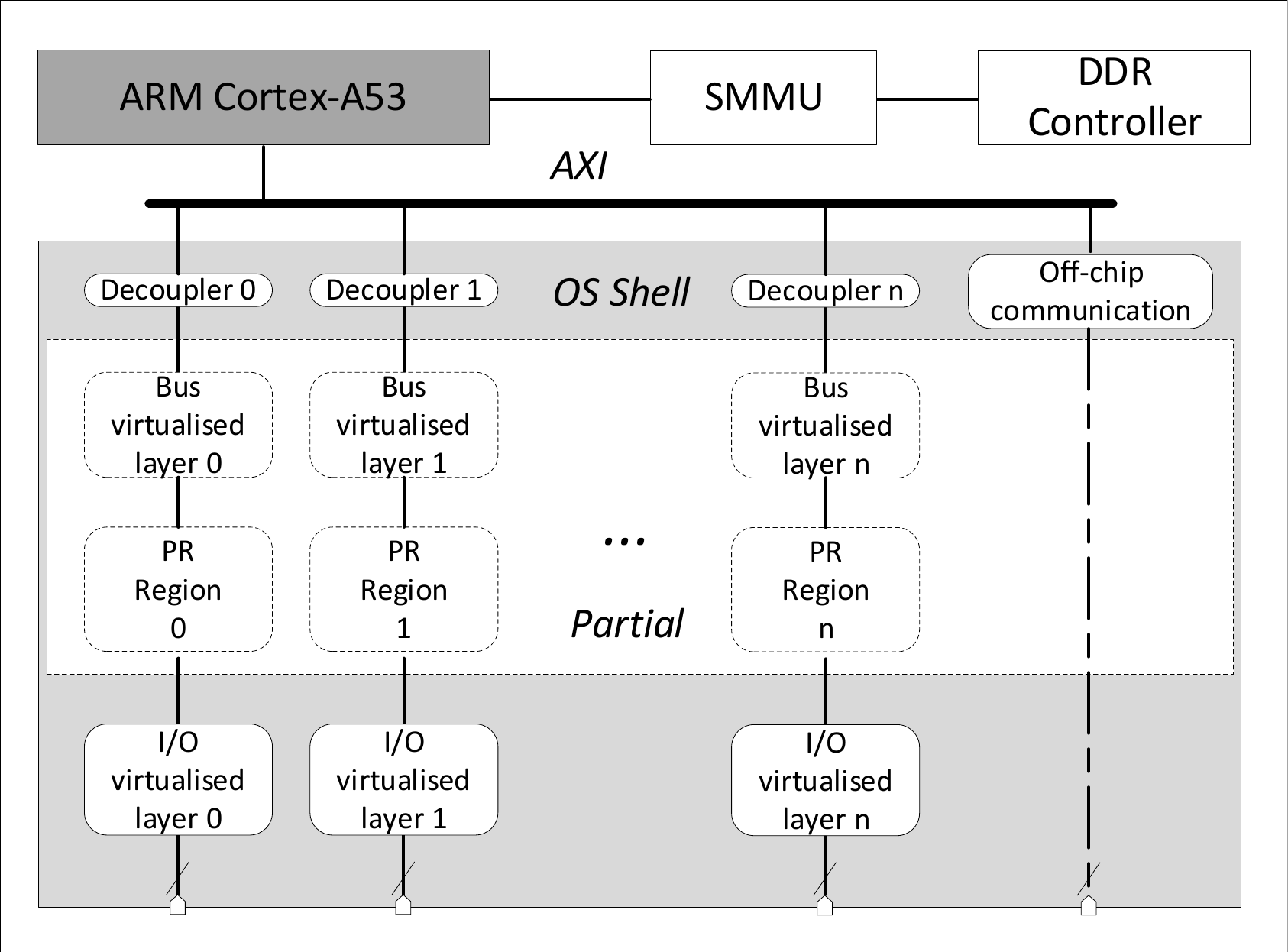}
	\caption{The overall organisation of an FPGA shell.}
	\label{fig-fos-overview}
\end{figure}

The hardware infrastructure is implemented directly on top of the physical FPGA resources and is commonly built using a partial reconfiguration (PR) flow (e.g., Amazon F1 FPGA instances~\cite{aws}). A PR flow generates a system in two partitions: a static system and one or more reconfigurable regions. An FPGA shell is essentially the static system of the PR flow and can be considered equivalent to the OS kernel to the hardware infrastructure. It provides common functionalities required by the accelerators such as an interconnect, network and memory access as well as (optional) other I/O and management IPs necessary for the target system. The counterpart of the shell is the partially reconfigurable region (also called slot) which can host different hardware accelerators at runtime. A shell which supports multiple partial regions can support multi-tenancy in the spatial domain by allowing multiple accelerators to execute in parallel. Figure~\ref{fig-fos-overview} shows an example of such a shell. However, the number of partial regions, their location and sizes depend on the system requirements and changes from system to system. Hence, an integral part of the shell is the compilation flow required to map the hardware modules onto the resources of a PR region with the required physical interface for the target system infrastructure. 

Moreover, a shell often includes a CPU in the form of a soft-core (e.g., on the datacenter FPGAs~\cite{xilinx_xrt}) or a hardened-core on MPSoC platforms used for embedded systems~\cite{xilinx_xrt}. This CPU is used to host the software infrastructure necessary for the management of the FPGA resources (see Section~\ref{sec:soft_infra} for details). 

With this, a shell provides the basic OS functionality at the hardware level to host hardware applications (accelerators) on an FPGA.


\subsubsection{Software Infrastructure}
\label{sec:soft_infra}

The software infrastructure of an FPGA operating system serves as a middle layer between the hardware accelerators and the user applications (software) while executing on a host CPU (soft or hardened). However, unlike software operating systems, the software infrastructure for an FPGA OS should be resilient to dynamic changes in the underlying hardware and be portable to different FPGA boards while hiding the heterogeneity from the software programmer. 
Consequently, an FPGA OS is responsible for managing the heterogeneity of the hardware resources available on the FPGA as well as to provide the high-level APIs and software integration for ease of development. This requires the software infrastructure to provide five important functionalities:
\begin{enumerate}
    \item A boot-loader and a kernel to bring up the FPGA system in an operational state.
    \item A set of drivers and hardware abstraction layer (HAL) to communicate with the hardware accelerators.
    \item A scheduler to dynamically allocate FPGA resources (i.e. partial regions) and hardware accelerators to software applications.
    \item High-level standard libraries to make hardware acceleration easily accessible. 
    \item Integration with existing CPU software stacks to benefit from legacy code.
\end{enumerate}


\subsection{Modular Development Flow}
\label{sec:modular_dev_flow}

\begin{figure}
    \centering
    \includegraphics[width=0.85\linewidth]{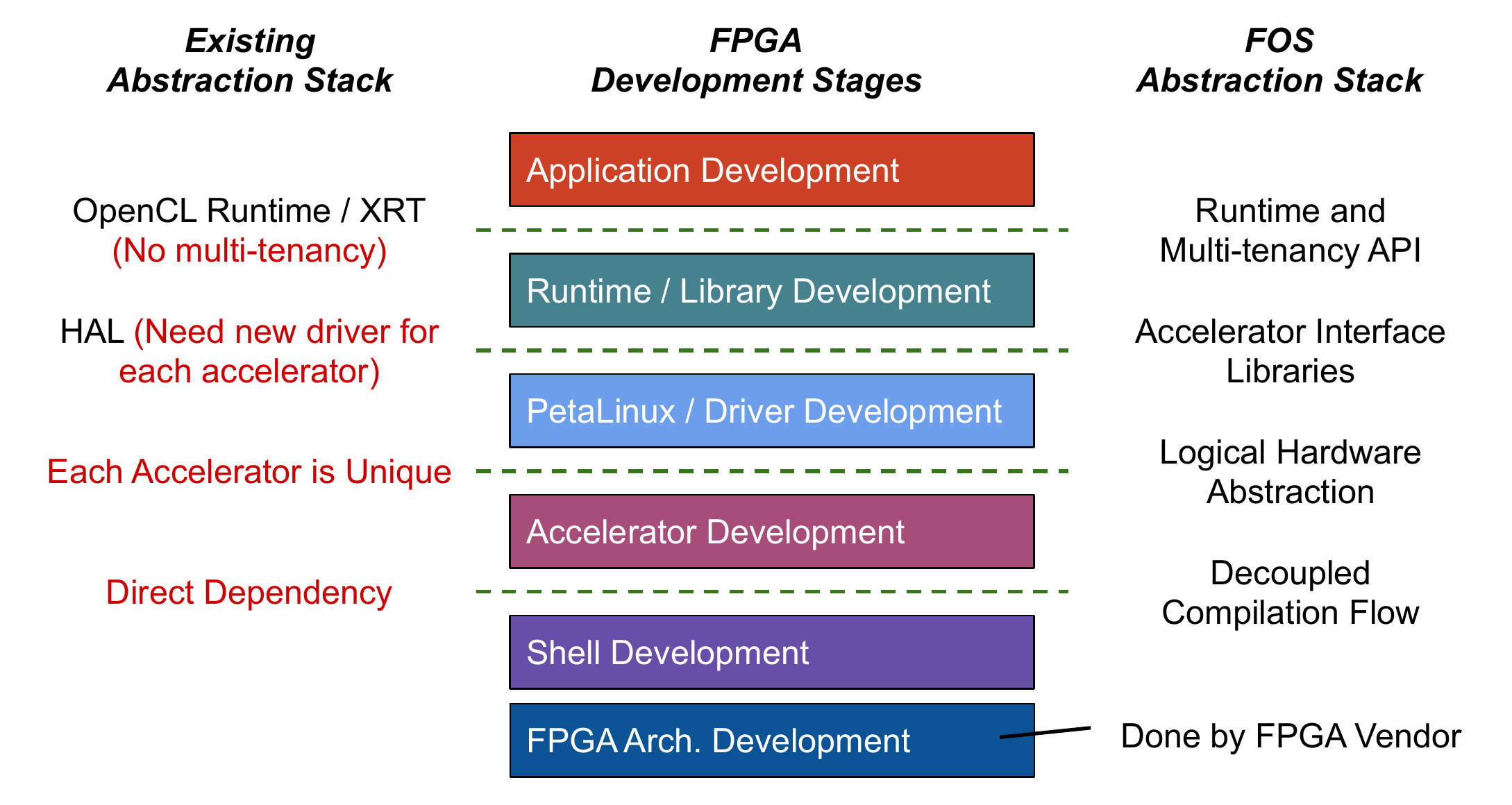}
    \caption{The abstraction layers required between FPGA development stages.}
    \label{fig:fos_abstraction_stack}
\end{figure}

There are five primary stages of development required for a multi-tenant/cloud FPGA system, as shown in Figure~\ref{fig:fos_abstraction_stack}. The bottom two stages of this stack are hardware development stages which drive the application performance and the support for multi-tenancy (via PR). In particular, 1) shell development requires designing and implementing the common system functionalities required by the user accelerators. An important step at this stage is floorplanning, which includes a decision on how many resources we allocate to the accelerators and a definition of the interface exposed to the accelerators. This stage also requires identifying the address mappings at which the driver layer will access the accelerators. While, its counterpart, 2) accelerator development, involves designing RTL or HLS accelerators with application-specific optimisations (commonly for the highest performance for the allocated resource budget). To compile the accelerators for the shell, the compiler must know the exact resources available in the partial regions and the physical locations of the interface pins. Without this information, we cannot implement partial reconfigurable accelerators. However, with the current development flow, accelerator compilation directly depends on the shell and requires knowing the implementation (internal resource allocation and routing) of a shell~\cite{xilinx2019partial}. This is because accelerator modules are implemented as an increment to a specific shell. Hence, any update made to the shell requires recompiling of the accelerators. Note that this flow is used to build the most state-of-the-art shells~\cite{khawaja2018armophos, ibmZurich, Fahmy, OliverKnodelVirtualizing, hcode}. 

The remaining three stages of the stack are software development stages which aim at programming the accelerators, runtime optimizations, and improving the developer productivity.
3)~PetaLinux/Driver development involves building the embedded software required to boot up the board with necessary I/O and high-level functionalities as well as means to communicate with the accelerators to send and receive data. The current Xilinx tool flow expects the user to write a new driver for each accelerator or generate one automatically when using Vivado HLS~\cite{feist2012vivado}. 
This new driver has to be either built with the embedded Linux kernel~\cite{petalinux} directly (for foreknown hardware) or with a device-tree overlay (for hardware known only at runtime). 
EDA tool vendors provide a hardware abstraction level (HAL) to make the development of the drivers easier via APIs to read and write to accelerators, to perform reconfiguration calls as well as basic C functionalities to enable debugging and fast development~\cite{pynq}. 
However, an accelerator developer or embedded Linux developer must take the responsibility to write and integrate the driver correctly in the rest of the Linux environment. 

The layer above this fundamental driver and kernel layer forms 4)~a runtime system or libraries for hardware acceleration. This provides a high-level API such as OpenCL to the user for serving reconfiguration and acceleration requests to the lower hardware layers~\cite{sdsoc,sdaccel, pynq}. However, currently, no FPGA vendor tools provide support for multi-tenancy.

Finally, 5)~in the application development stage a developer (commonly a domain expert) performs the required task in software while using hardware acceleration where appropriate. 
Note that a developer at this stage does not need to possess the skills required to implement the underlying FPGA development stack.

Overall, each step requires sophisticated knowledge which makes designing systems challenging for individual or small design teams. 
Given the complexity and the effort required for the task involved, often the final system is application-specific and cannot be reused or ported for different needs. To avoid this, we need a standard set of APIs such that a component can be swapped at each stage without recompilation or redevelopment of the components above or below it as long as the APIs are maintained. Thus, allowing each stage to be a modular artefact in itself. 

Such an abstraction stack would also allow 1) the software stack to be reusable across all types of FPGAs and FPGA boards, 2) not require the accelerator developers to write drivers, and 3) update system components in the shell with no need to propagate the changes to other stages in the stack. 

We can achieve this by implementing 4 key layers of abstractions: 
\begin{itemize}
    \item \textbf{Decoupled Compilation Flow}: Ability to compile accelerators and shells in isolation from each other. Whereby the accelerator compiles against a fixed physical interface and a bounded resource region. 
    This stops any changes in IPs or the shell from propagating to accelerators (given that we do not change the PR region and its interface).
    
    \item \textbf{Logical Hardware Abstraction}: Exposing the accelerator and the shell in a high-level format with only a minimum set of parameters required to build the drivers and perform the resource allocation. For the shell, this would include information such as how many regions it supports and what addresses they can be accessed at. Whereas, for the accelerators, it would include the i)~internal hardware address register mapping (ADR map) for programming the accelerator and ii)~meta-data associated with the accelerators for scheduling and management purposes (e.g., the size or maximum execution latency). 
    Note that HLS compilation (through Vivado HLS) provides this information completely without the need of any manual step.
    
    \item \textbf{Accelerator Interface Libraries}: 
    Using a standardised register mapping format to provide generic driver support for accelerators with streaming or master-slave interfaces (which are the most common interfaces provided by shells).
    Thus, it relieves the accelerator developer from the responsibility of writing drivers. In the case that the adoption of the standardised format is not feasible, a developer can use HAL APIs to build drivers.
    
    \item \textbf{Runtime and Multi-tenancy API}: Execution API which can abstract the interface at the logical level and which can span across multiple languages to provide high level integration to existing software stacks, while supporting the necessary primitives or programming models for dynamic resource allocation. 
\end{itemize}

\section{FOS Overview}
\label{sec:fos_overview}
The concepts for building an FPGA operating system (as introduced in Section~\ref{sec:fpga_os}) have been developed into FOS. 
FOS is a modular and lightweight FPGA Operating System which provides three primary modes of operation (as shown in Figure~\ref{fig:fos_usage_mode}): 
\begin{enumerate}
    \item Execution on static accelerators in a single-tenant mode. 
    \item Using multiple partially reconfigurable accelerators in a single-tenant mode. 
    \item Dynamically offloading acceleration request in a multi-tenant mode. 
\end{enumerate}

To provide the first two-modes of operation with ease of access, FOS provides two libraries: Cynq and Ponq. They provide high-level APIs to interact with hardware from C++ and Python applications, respectively. These APIs are platform-independent and can support the traditional Xilinx PR flow as well as the decoupled compilation flow introduced with FOS. 

The multi-tenant mode is provided through a daemon, which orchestrates the hardware acceleration requests from different users and schedules them in time \textit{and} in the spatial domain. Moreover, the daemon can execute hardware acceleration requests on heterogeneous accelerators (i.e. accelerators written in different languages) concurrently while providing the same user interface.

To support the underlying hardware interaction in each mode, the libraries include generic drivers to program accelerators and the Xilinx FPGA manager~\cite{petalinux} for partial reconfiguration. Moreover, FOS uses the PetaLinux Kernel~\cite{petalinux} and Ubuntu rootfs, to adopt the standard functionalities provided by the Ubuntu Linux distribution such as access to file systems, network access (via host-CPU), standard libraries, debugging tools, and other forms of I/O. 

The hardware infrastructure used in this paper is based on the open-source ZUCL 2.0 shell~\cite{zucl_2}. 
It currently supports three boards of varying FPGA capacity: ZCU102 (Xilinx MPSoC development kit), UltraZed, and Ultra-96 (suitable for IoT and edge deployment). The shell provides the ability to reallocate hardware accelerators across different partial regions as well as to combine multiple adjacent partial regions to host bigger accelerators. 

With these additions to the software and hardware ecosystem for FPGAs, FOS achieves a similar level of support for rapid development, deployment, and ease of use, as known from standard operating systems for CPUs. 

\section{Implementation}
\label{sec:implementation}
\subsection{Decoupled Compilation Flow for Shell and Modules}
\label{sec:decoupled_compilation_flow}
The primary requirement for abstracting accelerators from the shell is to decouple their compilation. However, this alone does not enable the flexible resource allocation required for the maximum utilisation of the resources. This is because of two main reasons, i)~the standard vendor EDA tool flow cannot easily combine partial regions to host bigger accelerators and ii)~it requires the accelerator to be compiled for each PR region it will be hosted in, hence resulting in multiple bitstreams for the same accelerator~\cite{xilinx2019partial}. To solve this, we need a shell that can support module relocation and re-adjustable PR regions (i.e. the ability to combine PR regions).
There are strict requirements to achieve development isolation with relocation and PR region flexibility: 
\begin{enumerate}
    \item PR regions should be homogeneous in terms of their resource foot-print (i.e. the relative layout of FPGA primitives) in order to allow for accelerator relocation, and regions should be located adjacent to each other to allow hosting of bigger accelerators without interfering with other system components.
    
    \item Communication interfaces between modules and the static system must be identical in terms of both logical protocol and their physical implementation such that relative positions of connection wires are the same in all PR regions. This ensures that modules can receive operation commands from the host CPU and it provides interfaces to transfer data back and forth to the main memory, irrespective of a module placement position.
    
    \item We must distribute clock signals in the same regular pattern across every PR region. These constrained clock routing paths will be used to provide clock signals to the resources used by the modules.
	\item We must prohibit routing from the static part to pass through the reconfiguration part and vice versa (except the interface) to ensure that module relocation does not interfere with other parts of the system.\footnote{This is not a strict requirement and the tool GoAhead~\cite{goahead} used to implement the shell can be used in a mode that allows static routing to cross reconfigurable regions.}
\end{enumerate}

\begin{figure}
	\centering
	\includegraphics[width=\linewidth,trim=15 15 15 15,clip]{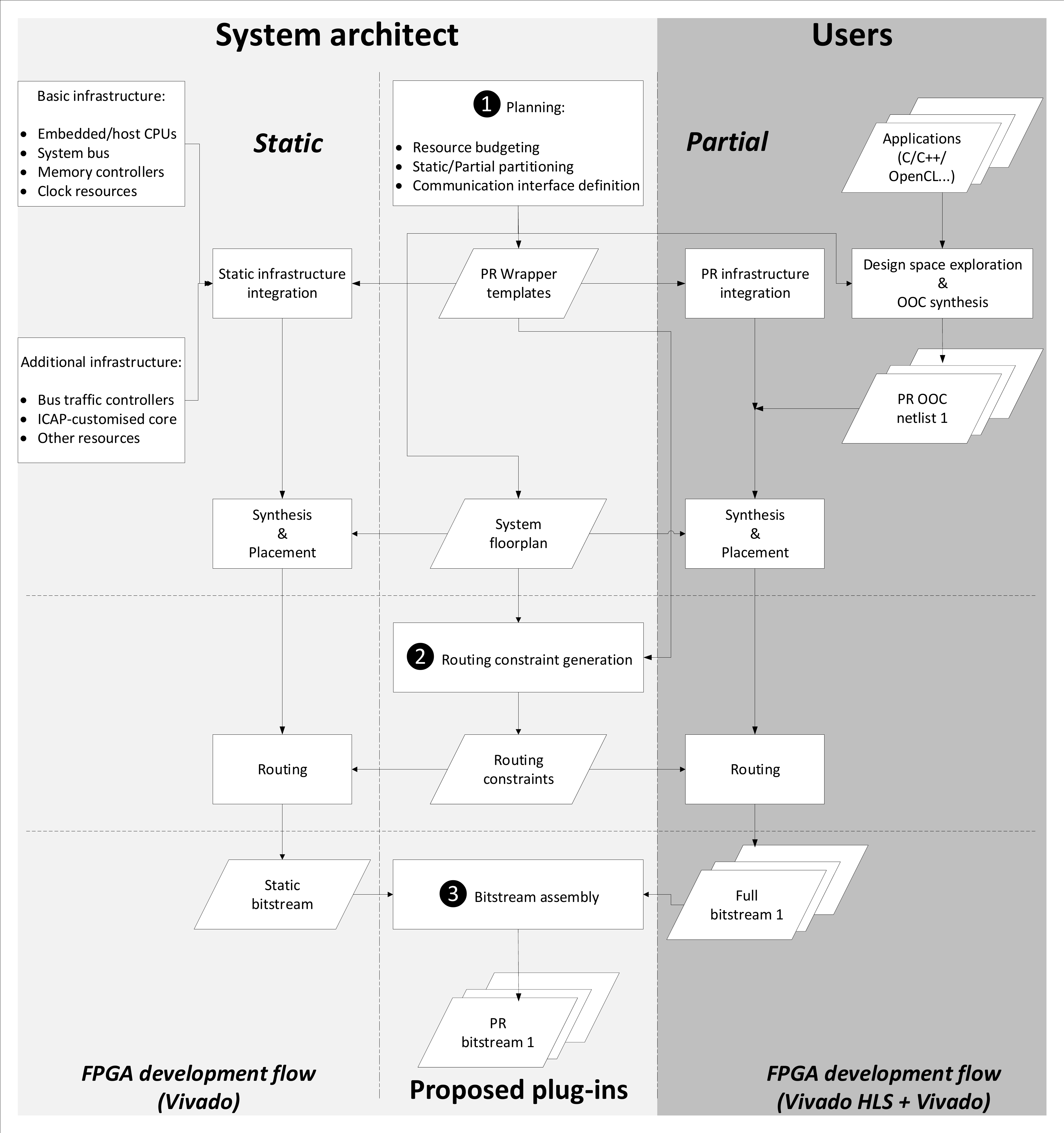}
	\caption{The proposed design methodology. The left part is performed once for a specific version of a shell by the system architect while the right part (with the darker background) is performed once for each module by FOS users.}
	\label{fig-flow-overall}
\end{figure}

The proposed design methodology depends on the standard FPGA development flow (Vivado toolchain for Xilinx FPGAs) to implement 1)~the basic infrastructure, acting as the OS shell, and 2)~the hardware applications (i.e. accelerator modules). Our contributions, however, include all the additional design steps and their corresponding tools that customise and adapt the default flow to automatically build the final system.
This integrates the entire shell and module development process into a unified design framework, as illustrated in Figure~\ref{fig-flow-overall}.

The main steps of the compilation process are as follows:
\begin{enumerate}
    \item \textit{Planning}: In this step, a shell developer needs to make a series of system-level design decisions.
	\begin{enumerate}
	    \item \textbf{Resource partitioning}: Based on the FPGA fabric layout and the number of resources available, a shell developer needs to split the FPGA fabric into two parts: 1)~the static region for FPGA shell and 2)~one or more reconfigurable regions for hosting hardware accelerators. 
	    This also defines the size of each PR region on which the High-Level Synthesis (HLS) tools~\cite{feist2012vivado} can perform the Design Space Exploration (DSE) for throughput optimisation~\cite{muslim2017dse}.
	    The trade-off at this stage requires the developer to allocate the maximum amount of resources to the reconfigurable part while leaving sufficient resources to the static part for shell functionalities and future upgrades. 
	    This trade-off is case-dependent and requires a thorough understanding of the target FPGA device and system requirements.
        However, this step is only performed once for a particular system. The result of this process is 1)~the static system floorplan and 2)~bounding boxes of the reconfigurable regions.
	    
	    
	    \item \textbf{Communication interface definition}: Select the protocol and data-width for communication between the static system and the partial regions based on the system requirements.
	\end{enumerate}

	\item \textit{Routing Constraint Generation}: The system floorplan and the PR interface template from the previous step are used to generate routing constraints. 
	We can describe the implementation rules with the help academic PR frameworks (GoAhead~\cite{goahead} and a TCL library~\cite{vesper2018tedtcl}) as TCL files for routing constraints automatically. 
	These TCL constraints will then guide the routing stage of Vivado. 

	\item \textit{Configuration Bitstream Generation}: The proposed flow results in \emph{static bitstreams} for both shell and module designs. To compose \emph{partial bitstreams} for accelerator modules from these bitstreams, we use the bitstream manipulation tool BitMan~\cite{bitman}, see Section~\ref{sec:module_compilation} for further details.
\end{enumerate}

\subsubsection{Shell Development}

The static design starts with integrating basic infrastructure and additional infrastructure to the unified top-level design. In our designs the basic infrastructure includes 64-bit ARM Cortex-A53 CPU cores, AXI4 interconnects, memory controllers, Xilinx PR Decouplers for disabling/enabling static and module communication, clock management tiles for tuning module frequencies, and PR Module Interfaces. 
We can add other resources such as memory management or network communication IPs if required.


\begin{figure}[!t]
    \begin{minipage}{0.49\linewidth}
    \centering
    \includegraphics[width=.95\linewidth]{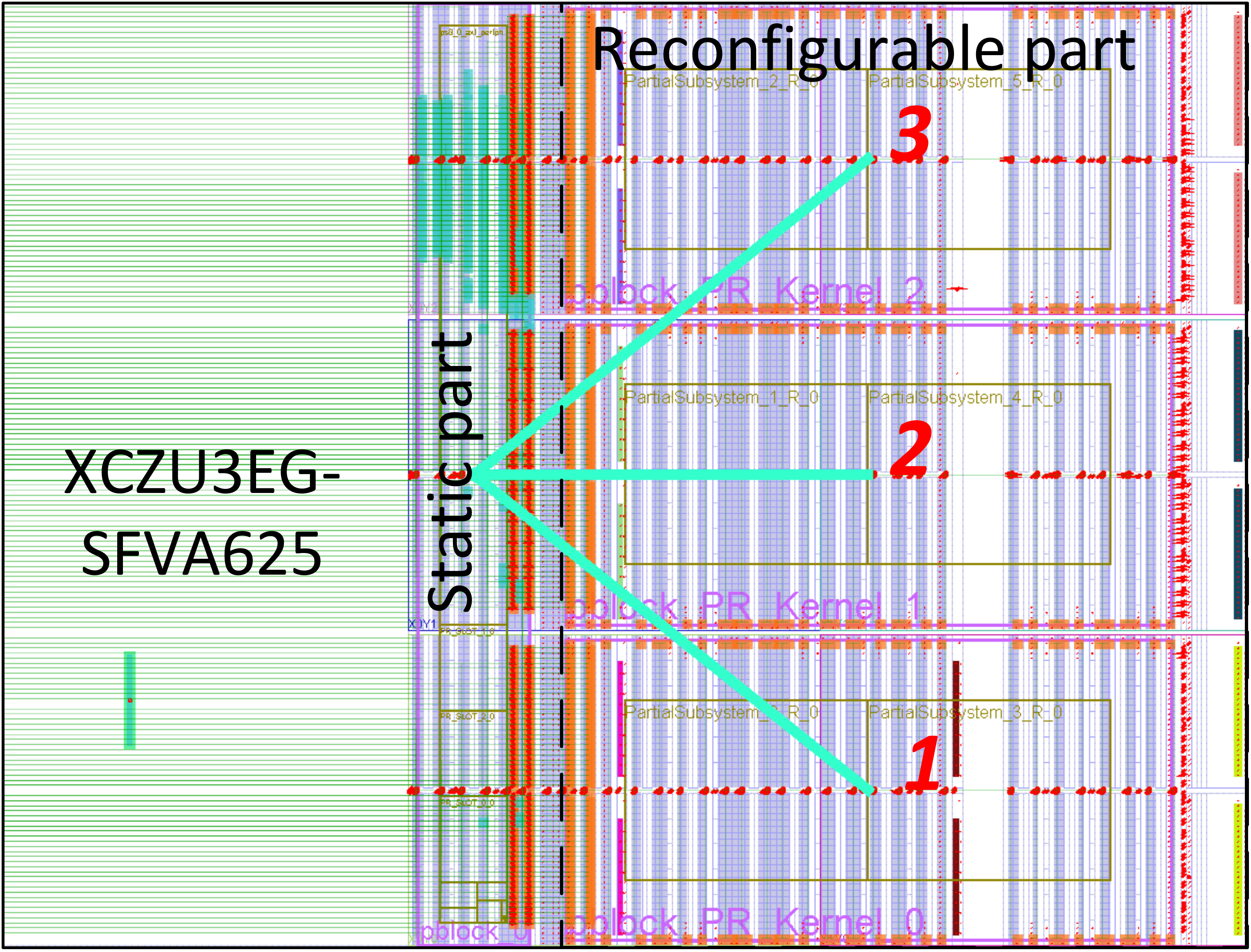}
    \caption{The physical shell implementation on the UltraZed and Ultra96 boards. This version has three PR regions which can host up to three FPGA applications simultaneously.}
    \label{fig-fos-impl}
    \end{minipage}
    \hfill
    \begin{minipage}{0.49\linewidth}
    \centering
    \includegraphics[width=0.75\linewidth]{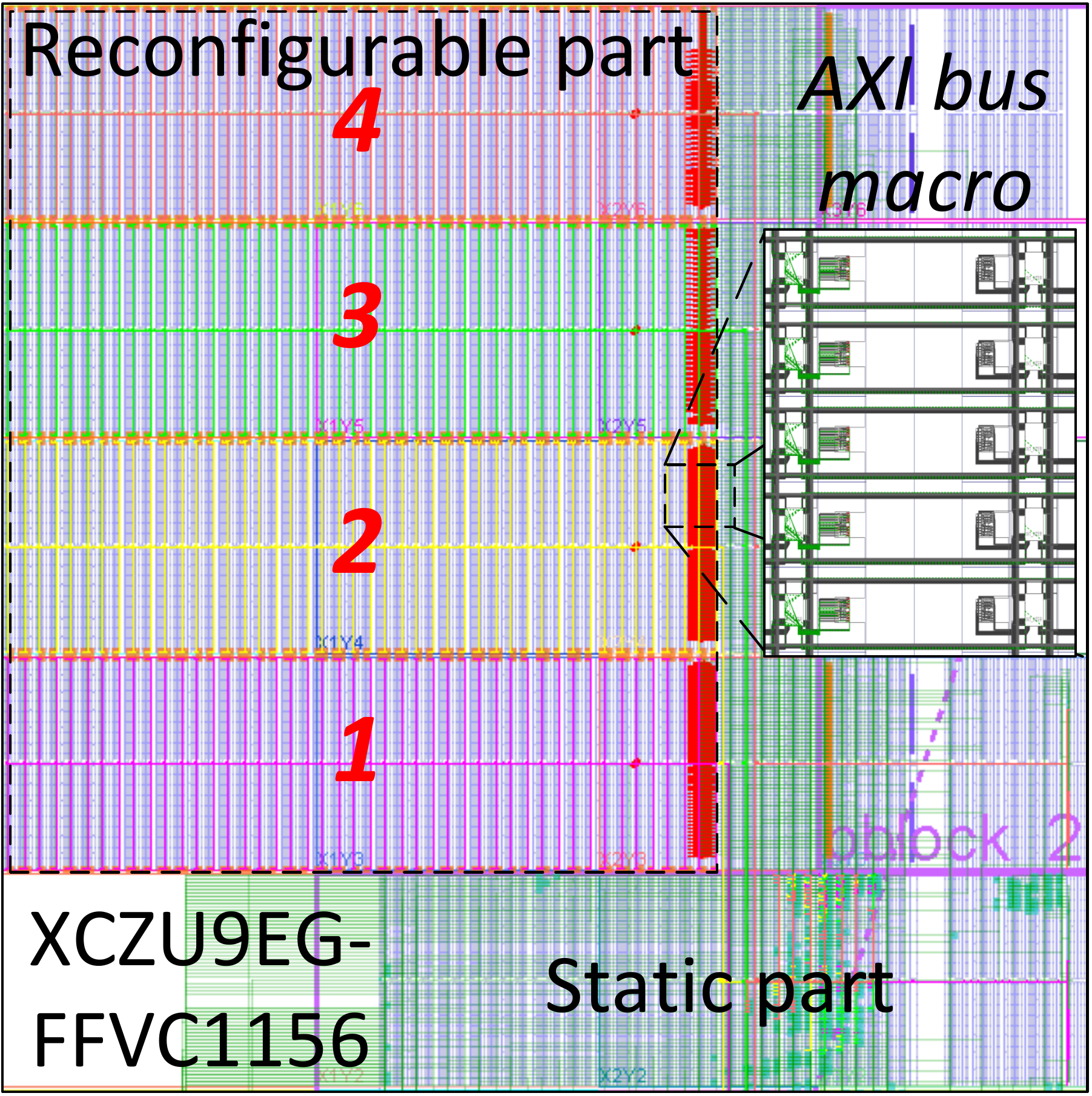}
    \caption{The physical shell implementation on the ZCU102 board. This version has four PR slots which can host up to four FPGA applications simultaneously.}
    \label{fig:zcu102_static}
    \end{minipage}
\end{figure}

The PR Module Interface provides an AXI4-Lite Slave for control register access via the CPU and an AXI4 Master for memory access.
This fixed interface between the hardware module and the static system is implemented using the available routing resources in the Zynq UltraScale+ FPGA devices.
In particular, we keep this interface identical for all PR regions to serve relocatable hardware modules by pre-placing and pre-routing these communication signals in a constrained, predefined manner. This reassembles the bus macro approach which was very popular for Xilinx Virtex-II devices~\cite{lysaght2006enhanced}. However, our flow can implement this without logic cost (in terms of LUTs used for the bus macro).

To keep the clocking resources of PR regions identical for relocatable hardware modules, we block all except for a defined subset of the \texttt{BUFCE\_LEAF} primitives inside PR regions.
This forces the router to use only a defined subset of these clock driver primitives that each drive a specific vertical clock spline which is ultimately connected to the flops, BRAMs, and DSPs in the region.
However, we only route the clock for the PR regions this way.
This means when routing the static system, 1)~we route the PR module clocks with \texttt{prohibit} constraints on the \texttt{BUFCE\_LEAF} primitives, then 2)~we remove these constraints and 3)~incrementally route the rest of the system.
This allows us to route additional clock nets as needed by the static system (e.g., for providing clocks to memory controllers or gigabit transceivers).


Finally, to prevent any static signal from violating PR regions, we insert a blocker macro. This blocker is non-functional, but uses all local wire resources inside the reconfigurable regions before routing the static system. We generate this blocker macro using GoAhead~\cite{goahead} or the TedTCL library~\cite{vesper2018tedtcl} according to the system floorplanning.

The above steps result in the final static design shown in Figure~\ref{fig-fos-impl} for the UltraZed/Ultra96 board and Figure~\ref{fig:zcu102_static} for the ZCU102 board.
We can see that the PR region interfaces have the same relative physical positions and the even distribution of clock splines across all PR regions.
Both systems support combining multiple adjacent regions for hosting larger monolithic modules. 
In this case, only one PR module interface will be used.

\subsubsection{Bus Virtualisation}
Operating a hardware accelerator needs communication with the host CPU to issue commands and to provide access to memory for data processing.
A module kernel can use a wide range of bus widths, such as 32/64/128-bit width, and various bus protocols, such as AXI4 Master/Stream in the proposed approach.
In particular HLS modules often, by default, include DMA engines for fetching data from memory and writing back results. However, this is not always the case with hand-crafted RTL or customised netlist accelerators. 

\begin{figure}[!t]
    \centering
    \includegraphics[width=.75\columnwidth,trim=15 15 15 15,clip]{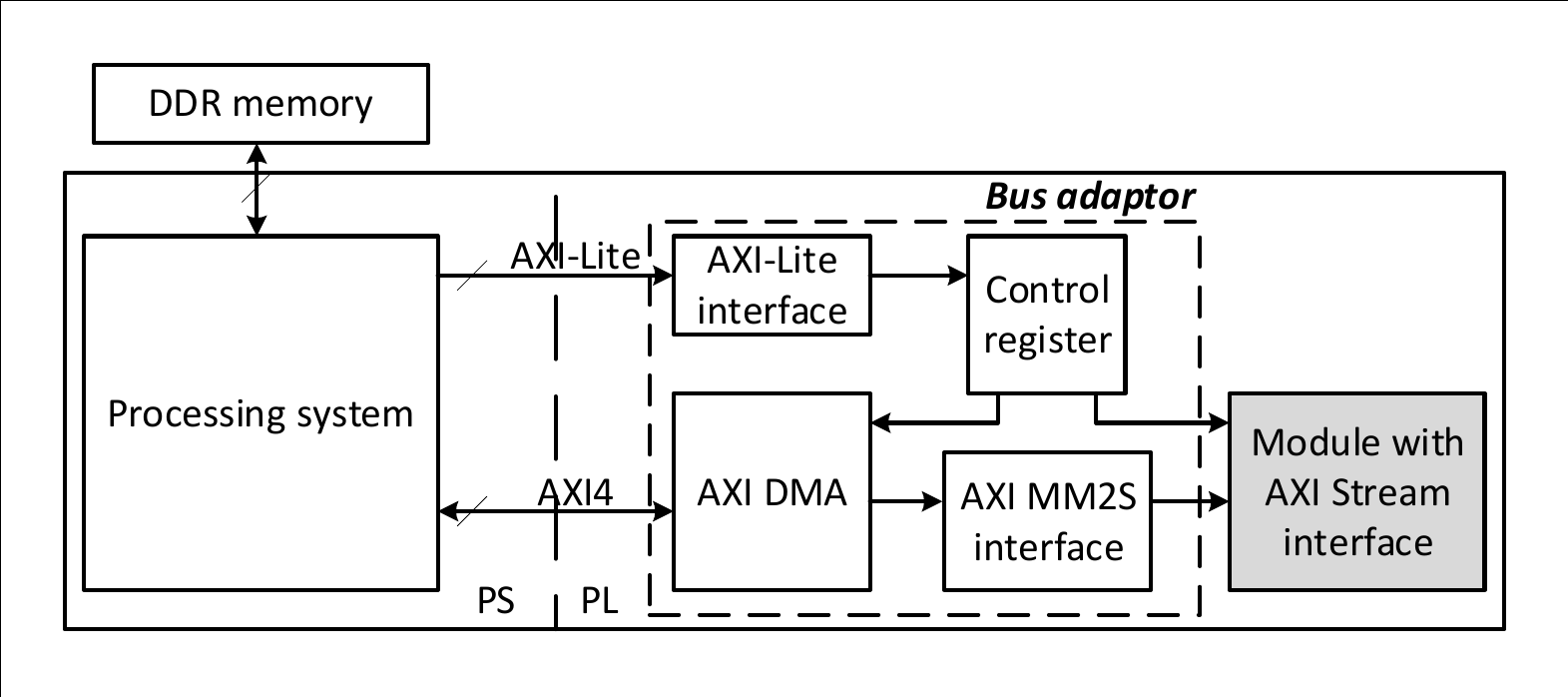}
    \caption{An example for bus virtualisation: the module has a 32-bit AXI-Lite interface and a 32-bit AXI Stream interface without DMA engine. In this case, the \emph{bus adaptor} with AXI DMA and AXI MM2S IPs are chosen to carry out the communication with the rest of system.}
    \label{fig-bus-virtualised}
\end{figure}

We tackle this issue by providing another level of abstraction for bus interfaces between the FPGA applications and shells. For this, we selected the interface to provide the 32-bit AXI-Lite protocol and the 128-bit AXI4 protocol.
Depending on the exact physical interface required by a module, we instantiate a module wrapper with a set of \emph{bus adaptors} such that a module can communicate with the rest of the system as required by the individual FPGA modules.
Figure~\ref{fig-bus-virtualised} shows a \emph{bus adaptor} being used to translate between different AXI bus standards. We can perform this wrapping at design time (where the wrapper is transparently instantiated when designing the module) or at runtime with partial reconfiguration (where the bus adaptor is a partially reconfigurable module located between the static system infrastructure and the accelerator module). The bus adaptor uses mostly IP components provided by FPGA vendor Xilinx~\cite{feist2012vivado}. These components are then automatically parameterised and integrated according to the specific interface requirements of the accelerator module. 
With this, shells can remain light-weight, operational, and unchanged while supporting a wide range of AXI interfaces. 

The present FOS shells support up to 128-bit wide datapath for memory accesses because this is the native width to the ARM SoC.
It is important to understand that a static acceleration system would also use such bus adaptors provided by the vendor, hence, our bus adaptors do not necessarily cause an additional overhead.
The advantage of the here proposed \emph{bus adaptor} concept is that an adaptor is only integrated into a module if needed and not speculatively provided by the shell.

\begin{figure}[!t]
    \centering
    \includegraphics[width=.72\columnwidth,trim=15 15 15 15,clip]{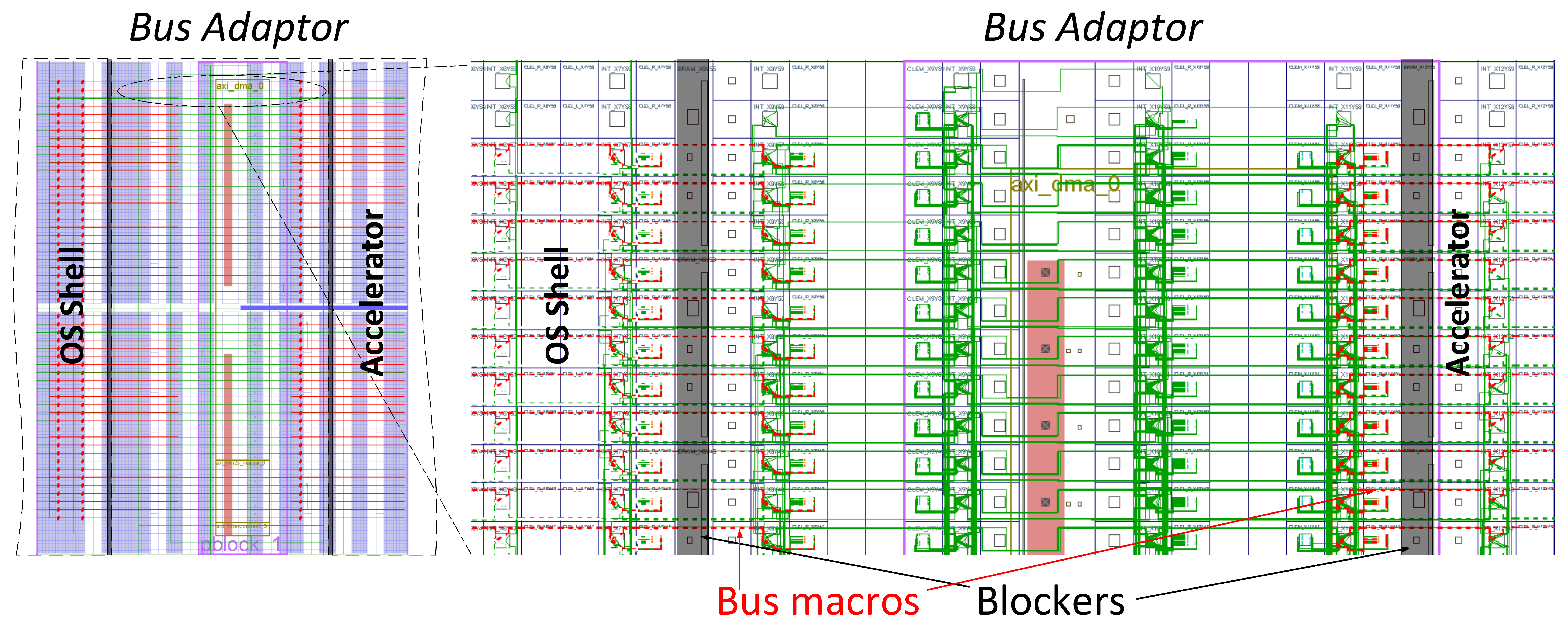}
    \caption{Implementation of a bus abstraction layer on UltraZed/Ultra96 platforms. The bus adaptor is provided as an implemented module bitstream and stitched to into the shell at run-time by using partial reconfiguration. The adaptor is a partial module that, in turn, interfaces to other partial modules. This technique avoids re-compiling bus adaptors but comprises an area overhead for a partial region to host a bus adaptor.}
    \label{fig-bus-virt-impl}
\end{figure}

We also use a bus adaptor to translate between AXI Master and AXI Stream protocols. 
We provide different versions of AXI Stream adaptors to be used depending on the AXI Stream channel width.
A user can either re-compile their modules with a logical wrapper of the bus adaptor at design phase or stitch their modules with a pre-built binary of that bus adaptor at run-time.
Figure~\ref{fig-bus-virt-impl} shows the implementation of the bus adaptor with its logic for interfacing an accelerator with the AXI protocol, AXI MM2S, and AXI DMA services for a module which has a 32-bit AXI-Lite and 32-bit AXI Stream interface.

\subsubsection{Module Compilation}
\label{sec:module_compilation}
The module design begins from either high-level language (HLL) source code (C, C++, OpenCL etc.) or a hardware description (RTL/netlist).
In the case of HLL source code, we go through a High-Level Synthesis (HLS) step~\cite{feist2012vivado} to generate the RTL source codes. 
We then synthesise the resulting RTL source code in out-of-context (OOC) mode.

The PR Wrapper templates are used to create a minimal top-level placeholder for the module implementation.
This temporary placeholder acts as sink/source connection points and substitutes the surrounding static system.
We then integrate the module OOC netlist into this placeholder for the synthesis and placement stages.

We generate blockers (as TCL routing constraints) to enforce that all partial module’s primitives and routing resources are following the strict implementation rules mentioned in Section~\ref{sec:decoupled_compilation_flow} (by using GoAhead~\cite{goahead} or TedTCL library~\cite{vesper2018tedtcl}).
We place these blockers around the selected area to act as a fence for implementing hard module bounding box constraints.
The blockers include routing tunnels for the communication to and from the temporary placeholder.
The position of these tunnels matches exactly the tunnels used in the static design to implement the communication between static and partial areas.

\begin{figure}
    \centering
    \begin{minipage}{0.48\textwidth}
    \includegraphics[width=\linewidth]{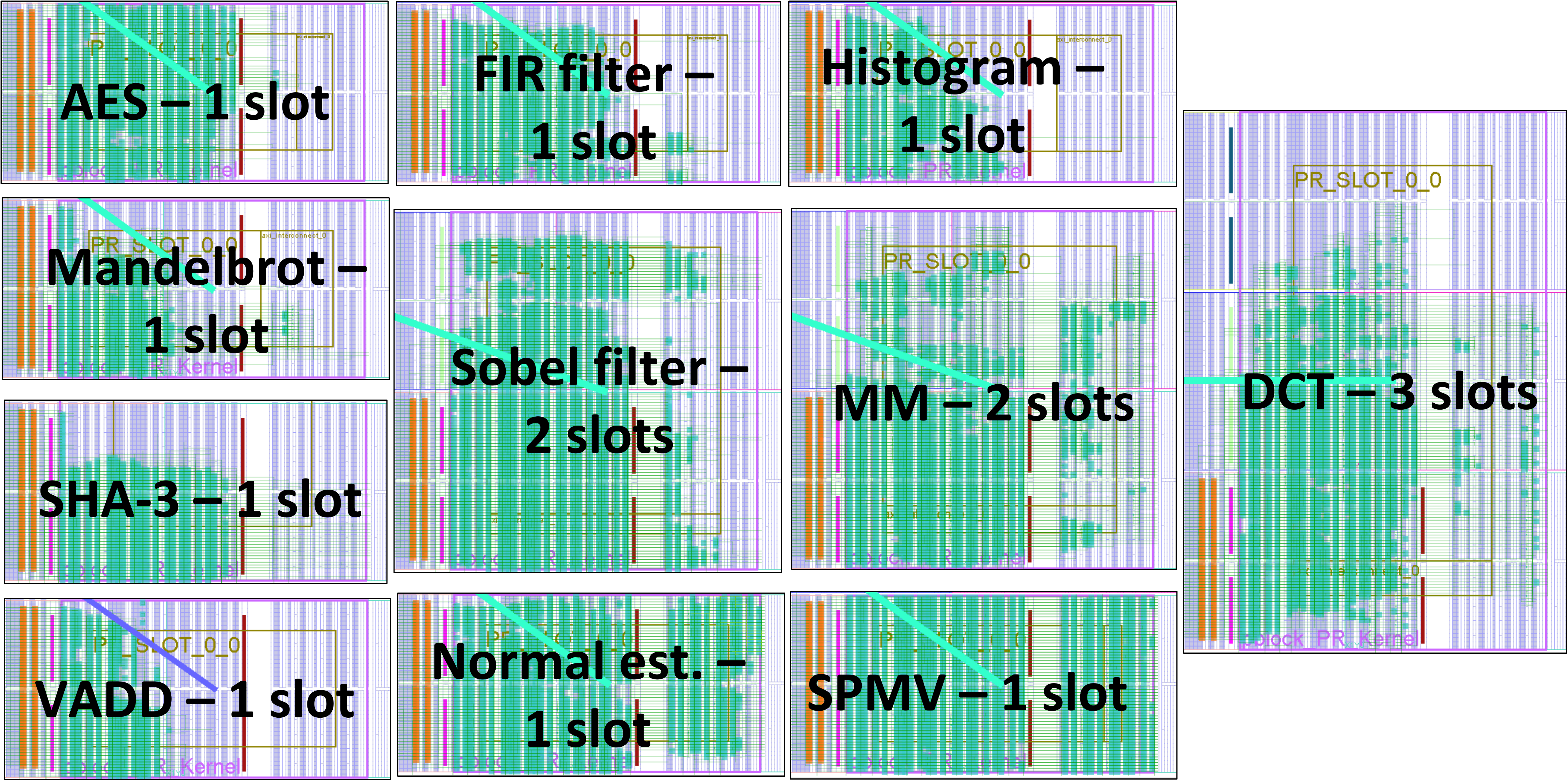}
    \caption{Compiled modules for Spector benchmark suite~\cite{spector} and our in-house accelerators for Ultra-96 board.}
    \label{fig:u96_modules}
    \end{minipage}
    \hfill
    \begin{minipage}{0.47\textwidth}
    \includegraphics[width=\linewidth]{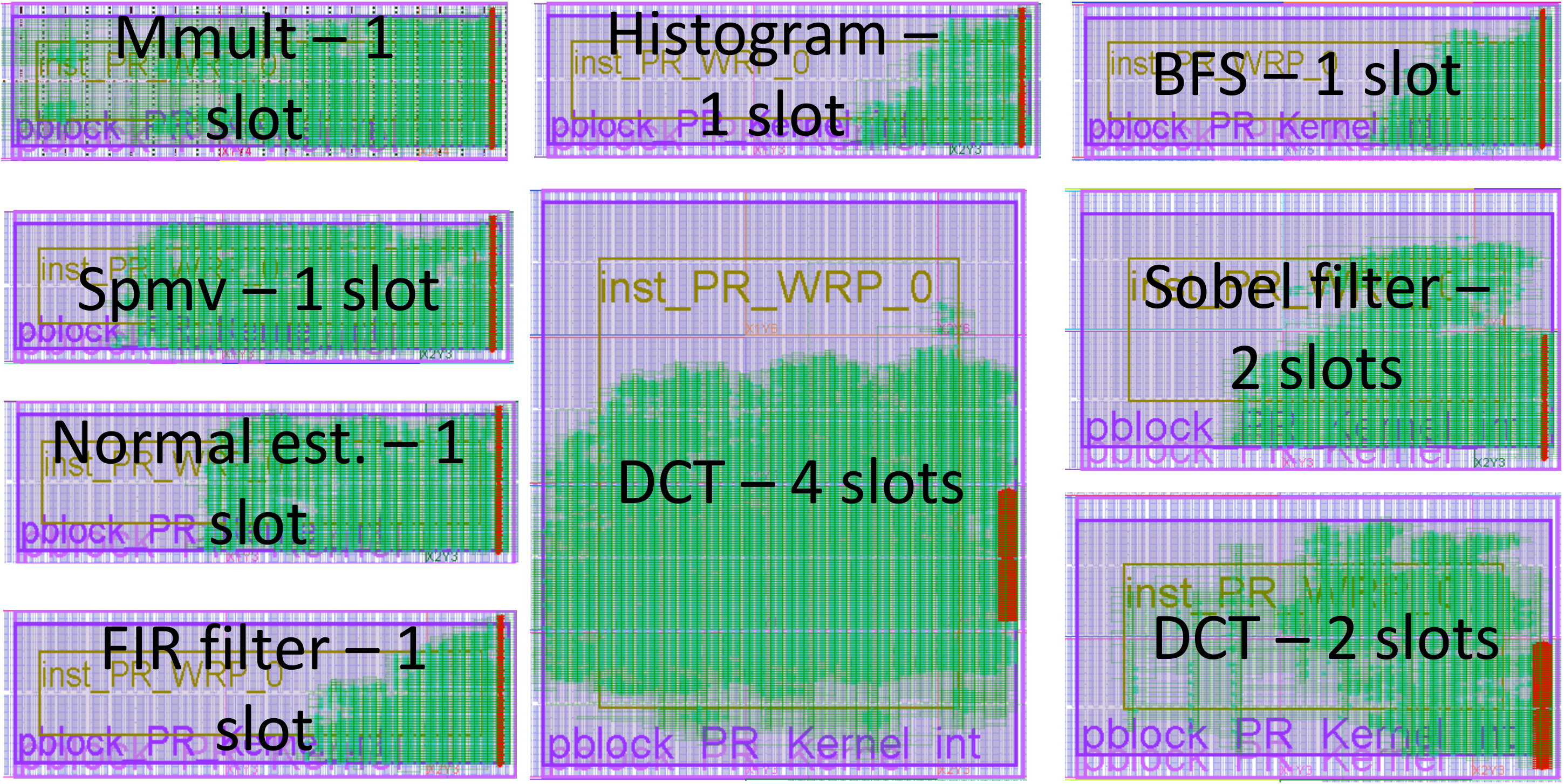}
    \caption{Compiled modules for Spector benchmark suite~\cite{spector} accelerators for ZCU102 board.}
    \label{fig:zcu102_modules}
    \end{minipage}
\end{figure}

As we implement a module in separation from the static system, the result generated by Vivado is a full configuration bitstream.
We pass this full bitstream to BitMan~\cite{bitman} to extract the configuration data that corresponds to the module only as a partial bitstream.
At run-time, BitMan manipulates those partial bitstreams to relocate modules to the desired partial region of the static system. Figure~\ref{fig:u96_modules} and~\ref{fig:zcu102_modules} shows the resulting modules for the Spector benchmark suite~\cite{spector} and our in-house accelerators for Ultra-96 and ZCU102 boards, respectively. 

\subsection{Logical Hardware Abstraction}
This is the primary layer between the hardware and software infrastructure. It is designed to hide the differences or changes in the hardware from the software, in order to detach the software infrastructure from the underlying hardware layer as much as possible. To achieve this conveniently, we propose describing the shell in terms of logical functionalities using a JSON file description with the following information (see listing~\ref{lst:shell_descript} for an example): 

\begin{enumerate}
    \item Name of the shell
    \item Bitstream name of the shell
    \item Partial region:
        \begin{enumerate}
            \item Name of the partial region 
            \item Blanking bitstream for the partial region 
            \item AXI bridge decoupler address
            \item Base address of an accelerator placed in the region
        \end{enumerate}
\end{enumerate}

\begin{lstlisting}[language=json, label={lst:shell_descript}, caption={JSON description example of a shell.}]
{
  "name": "Ultra96_100MHz_2",
  "bitfile": "Ultra96_100MHz_2.bin",
  "regions": [
    {"name": "pr0", "blank": "Blanking_slot_0.bin", "bridge": "0xa0010000", "addr": "0xa0000000"},
    {"name": "pr1", "blank": "Blanking_slot_1.bin", "bridge": "0xa0020000", "addr": "0xa0001000"},
    {"name": "pr2", "blank": "Blanking_slot_2.bin", "bridge": "0xa0030000", "addr": "0xa0002000"}
  ]
}
\end{lstlisting}

Similarly for the accelerator, we defined a JSON file description (see Listing~\ref{lst:accel_descript}) with the following details for accelerator programming and management:

\begin{enumerate}
    \item Name of the accelerator
    \item Bitstreams:
        \begin{enumerate}
            \item Name of the bitstream
            \item Name of the shell it is compiled for
            \item Type of AXI interface used (if not AXI4 master and slave)
            \item Name and number of the PR regions it is compiled for
        \end{enumerate}
    \item Register mappings:
        \begin{enumerate}
            \item Name of the HW register
            \item Address offset at which the HW register can be access
        \end{enumerate}
\end{enumerate}

Given that the control register map follows the standard Vivado HLS~\cite{feist2012vivado} interface (see Listing~\ref{lst:accel_control}), an accelerator can have an arbitrary number of 32-bit registers. This allows us to build generic drivers for accelerators to relieve hardware developers from the responsibility of writing and integrating drivers. 
Hence, with this logical hardware abstraction, we can arbitrarily update the shell or accelerators with no need to recompile the Linux kernel or drivers.
Section~\ref{sec:accel_interface_lib} will describe the driver and acceleration library interface.

The name of the PR region for each bitstream allows backward compatibility to the Xilinx PR flow, where we must compile an accelerator for each region (no relocation support). 
Moreover, the name of the accelerator is unique, but it may contain bitstreams of varying sizes corresponding to different acceleration implementations with the same functionality. The scheduler can later use these implementation alternatives to perform resource-elastic scheduling, i.e. dynamically change the resource allocation used by the accelerator based on the workload and available resources.

We then register these JSON descriptions for shell and accelerators into a JSON based registry to enable a centralised view of the available hardware to the upper software layers.
This allows the application developers or the runtime system to request hardware based on just the name (a logical accelerator functionality) and corresponding input data, without needing any further information about the underlying hardware layer and accelerator implementation.

Note that we can automatically generate the JSON descriptor for the accelerator from the files generated by the Vivado HLS compilation flow~\cite{feist2012vivado}. Whereas, the shell developer must write the JSON description to allow the runtime system to manage the resource allocation and use generic drivers. However, this process is commonly required only once per system and can be omitted entirely when using our pre-built FOS shells.

\begin{minipage}{0.48\linewidth}
\centering
\begin{lstlisting}[language=json, label={lst:accel_descript}, caption={JSON description of a vector add accelerator.}]
{
  "name": "vadd",
  "bitfiles": [
    {"name": "vadd.bin", "shell": "Ultra96", 
     "region": ["pr0", "pr1"]},
  ],
  "registers": [
    {"name": "control", "offset": "0"},
    {"name": "a_op",  "offset": "0x10"},
    {"name": "b_op",  "offset": "0x18"},
    {"name": "c_out", "offset": "0x20"},
  ]
}
\end{lstlisting}
\end{minipage}%
\hfill
\begin{minipage}{0.45\linewidth}
\centering
\begin{lstlisting}[language=C++, label={lst:accel_control}, caption={Control bits for the accelerator.}]
// 0x00 : Control signals
//        bit 0  - ap_start (Read/Write/COH)
//        bit 1  - ap_done (Read/COR)
//        bit 2  - ap_idle (Read)
//        bit 3  - ap_ready (Read)
//        bit 7  - auto_restart (Read/Write)
//        others - reserved
\end{lstlisting}
\end{minipage}%

\subsection{Acceleration Interface Libraries}
\label{sec:accel_interface_lib}

\begin{figure}
    \centering
    \includegraphics[width=0.48\linewidth]{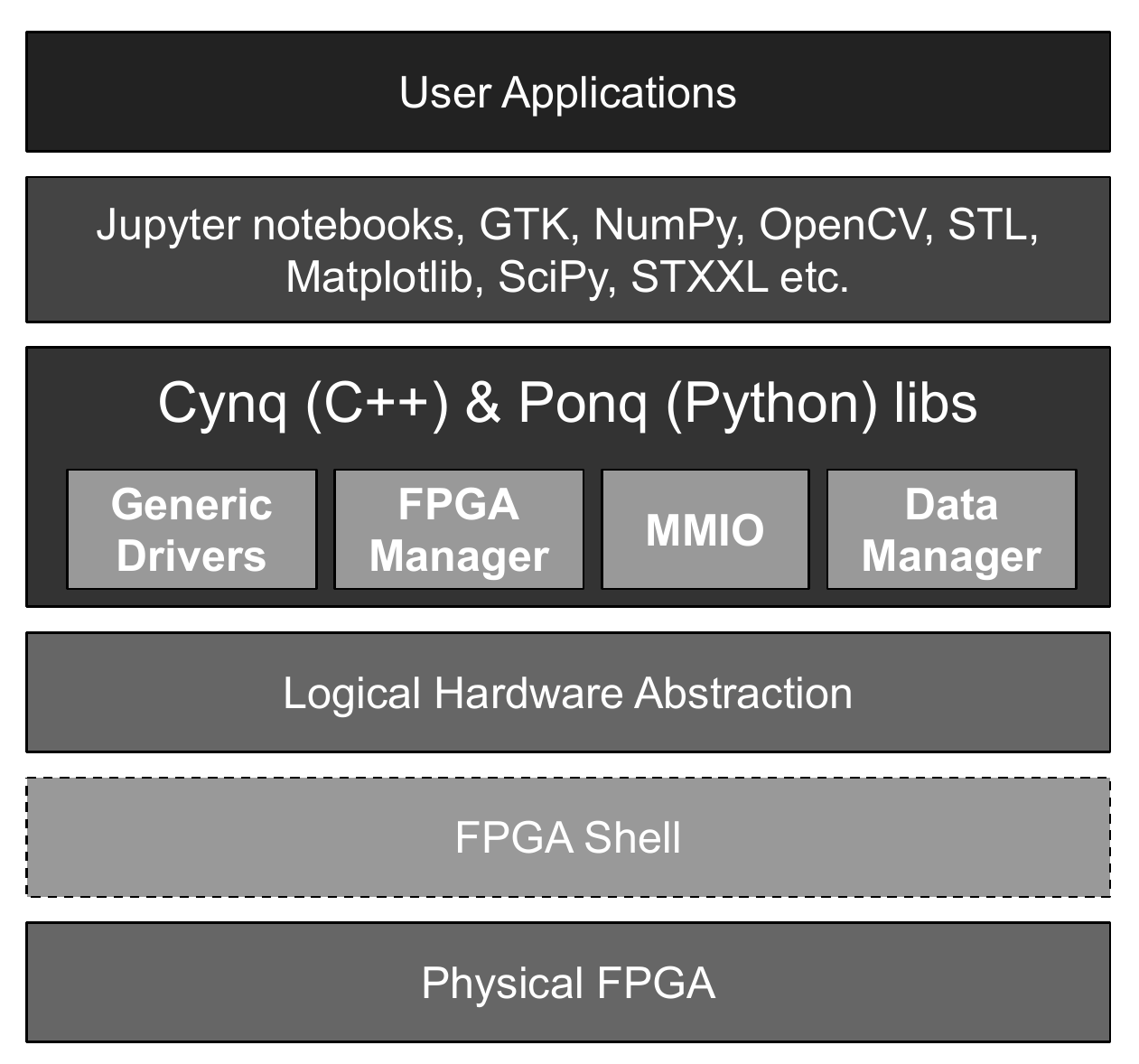}
    \caption{Cynq and Ponq libraries as an acceleration interface layer for static and dynamic acceleration on FPGAs.}
    \label{fig:accel_libs}
\end{figure}

In general, an OS has to provide high-level APIs that can be used from existing software stacks to improve the accessibility of FPGAs to software developers (who are often non-FPGA experts).
One such effort is the PYNQ~\cite{pynq} framework from Xilinx, which allows accessing FPGA accelerators through high-level Python APIs.
However, the current implementation of PYNQ relies on a traditional development flow and contains many direct dependencies on artefacts produced by the Vivado compilation flow which restrict modular development~\cite{pynq, feist2012vivado}. While it is possible to engineer around these shortcomings, the resulting system would suffer from code inflation and would be non-scalable due to its legacy support.
Hence, we built a new light-weight acceleration interface libraries called Ponq and Cynq for Python and C++ languages, respectively.
These libraries are built based on a modular development flow and provide access to static and dynamic FPGA accelerators via its generic drivers for programming accelerators, the Xilinx FPGA manager for partial reconfiguration~\cite{petalinux}, memory mapped I/O (MMIO) modules for direct access to accelerators and a data manager for contiguous physical memory allocation. 
Figure~\ref{fig:accel_libs} shows the integration of Ponq and Cynq libraries into the FOS modular development flow and existing high-level software libraries. 

Moreover, the libraries are backwards compatible with the standard Xilinx development flow, i.e. both the PR flow and the static acceleration environment, as we build them on top of the logical hardware abstraction layer. The libraries provide the following basic HAL functionality and generic drivers for hardware acceleration:
\begin{itemize}
    \item Load an FPGA shell
    \item Load a partially reconfigurable accelerator
    \item Load a static accelerator
    \item Load and program an accelerator based on a logical function name
    \item Program an accelerator for execution via generic drivers
    \item Read and write calls to HW registers of the accelerators
    \item Contiguous physical memory allocation
\end{itemize}

\subsection{Runtime and Multi-tenancy API}
\label{sec:fos_runtime}

To support multi-tenancy, a runtime system is necessary to arbitrate access to reconfigurable resources between multiple users \textit{transparently}.
The common approach for allocation is to employ time-domain multiplexing with a run-to-completion model, but there also exists a few spatial domain scheduling mechanisms for FPGAs~\cite{asiatici, resource_elasticity}. 
These spatial domain schedulers, however, only support a specific type of accelerator, such as OpenCL or DSL based accelerators.
An operating system, in contrast, must support all types of accelerators and allow them to execute concurrently while using space-time domain scheduling.
To make this possible, we need three main components: an API to a daemon, a programming model and a scheduler, as discussed in the following paragraphs. 

\subsubsection{Daemon API}

To truly support multi-tenancy with portability, we need to design an API which can span across multiple languages and be portable to different OS kernels or a base processor system with ease.
The two efficient ways to perform this Inter-process communication (IPC) are 1)~message passing and 2)~shared memory. 
In our platform we adopt the gRPC framework~\cite{grpc} which is a standard RPC framework with support for multi-languages. We use gRPC to send the acceleration requests from the client process to the daemon process, while the data is passed via shared memory to avoid additional latency of copying data (i.e. zero copy operation). 
The adoption of gRPC allows us to extend the runtime to accept acceleration requests from remote nodes in the future.
The final interface exposed to the application developer in C++ and Python is shown in Listing~\ref{lst:exec_c} and~\ref{lst:exec_python}.
Note that each user can offload multiple data-parallel acceleration requests in a single RPC call to the daemon. 

\begin{minipage}{0.47\linewidth}
\centering
\begin{lstlisting}[language=c++, label={lst:exec_c}, caption={C++ daemon execution call example.}]
// Create a job
Job &job = jobs.emplace_back();
job.accname = "Partial_accel_vadd";

// Set accelerator parameters
job.params["a_op"]  = a_op_phy_addr;
job.params["b_op"]  = b_op_phy_addr;
job.params["c_out"] = c_op_phy_addr;

// Launch jobs
fpgaRpc.Run(job);
\end{lstlisting}
\end{minipage}%
\hfill
\begin{minipage}{0.47\linewidth}
\centering
\begin{lstlisting}[language=python, label={lst:exec_python}, caption={Python daemon execution call example.}]
# Create a job and set accelerator parameters 
jobs = [{
    "name": "Partial_accel_vadd",
    "params": {
        "a_op":  a_op_phy_addr,
        "b_op":  b_op_phy_addr,
        "c_out": c_op_phy_addr,
    }]

# set accel parameters and run hardware unit
fpga_rpc.Run(jobs)
\end{lstlisting}
\end{minipage}%

\subsubsection{Programming Model}

We achieve the dynamic resource allocation in the spatial-domain using resource-elasticity~\cite{resource_elasticity} in two forms: module replication and module replacement. 
However, to make this possible for all types of accelerators we allow applications to expose data-parallelism to the scheduler, i.e. an application developer can choose to express an acceleration job into a varying degree of parallelism appropriate for the application. 
A common example of this is chopping the image into multiple parts for image-processing accelerators.
Note that this is analogous to a software developer deciding for the number of threads for efficient execution on a multi-core system.  
The runtime is then responsible for executing those parts i)~in parallel, ii)~use a better implementation, or iii)~in case of a exceeding FPGA resource capacity perform time-domain multiplexing for the resources. 

\subsubsection{Scheduling}

\begin{figure}
    \centering
    \includegraphics[width=0.90\linewidth]{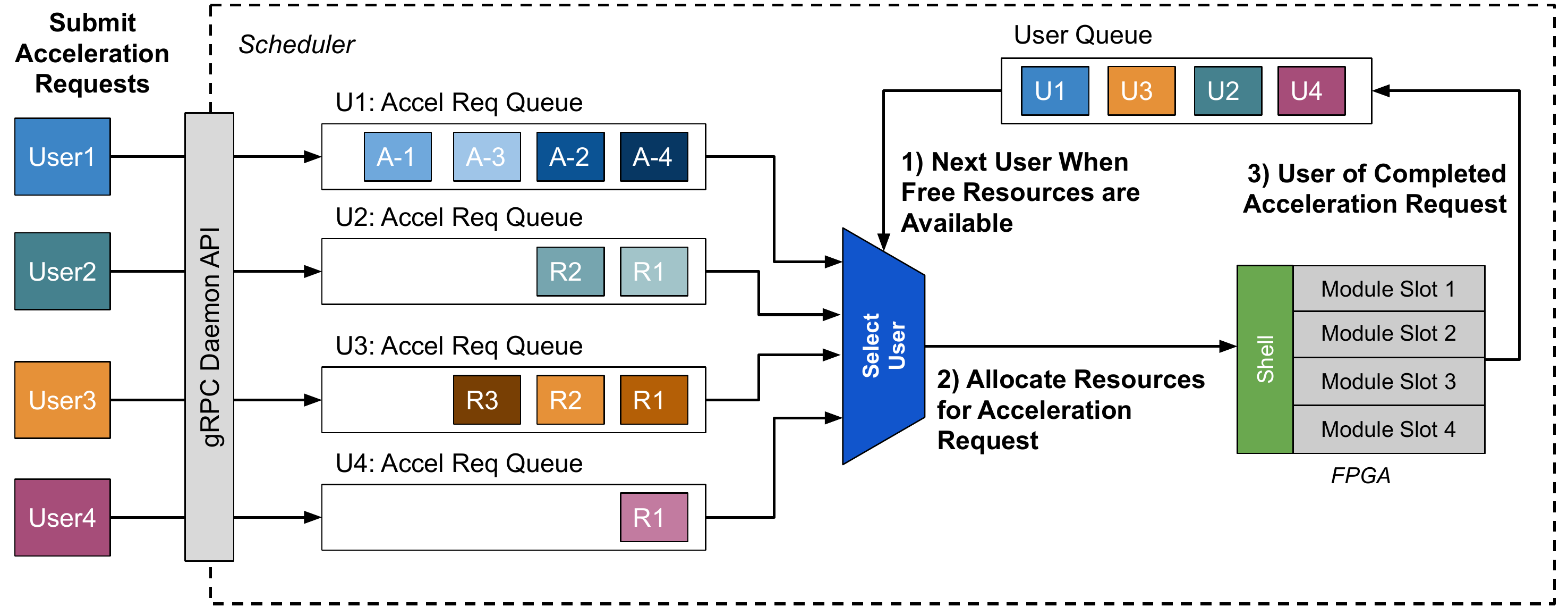}
    \caption{Scheduler organisation for resource allocation between different users and their data-parallel acceleration requests.}
    \label{fig:fos_scheduler}
\end{figure}

The scheduler maintains a queue of the users and performs round-robin scheduling between users at a coarse granularity of data-parallel acceleration requests, as shown in Figure~\ref{fig:fos_scheduler}. Each user has an individual queue of acceleration requests. Each request in this queue is independent of other requests in the queue and can execute in parallel and any order. At the end of each acceleration request, the scheduler relinquishes the accelerator and selects an acceleration request from the next user in the queue. 
This scheme implements a cooperative scheduling policy (by breaking down a job into fine-grain run-to-completion acceleration requests) where each request includes fetching operands and writing back results to main (DDR) memory. This corresponds directly to the OpenCL programming model where work-groups can be executed in any order.

In the case there is no other user, the scheduler executes requests from the same user in parallel and attempts to use the biggest module (assumed to be the fastest, i.e. Pareto-optimal) to maximise the utilisation and performance.
Moreover, the scheduler avoids partial reconfiguration and reuses an accelerator if it is already available on-chip. This allows multiple different applications that require acceleration of the same functionality to share an accelerator in time without paying an additional penalty or user effort. 

\begin{figure}
    \centering
    \includegraphics[width=0.75\linewidth]{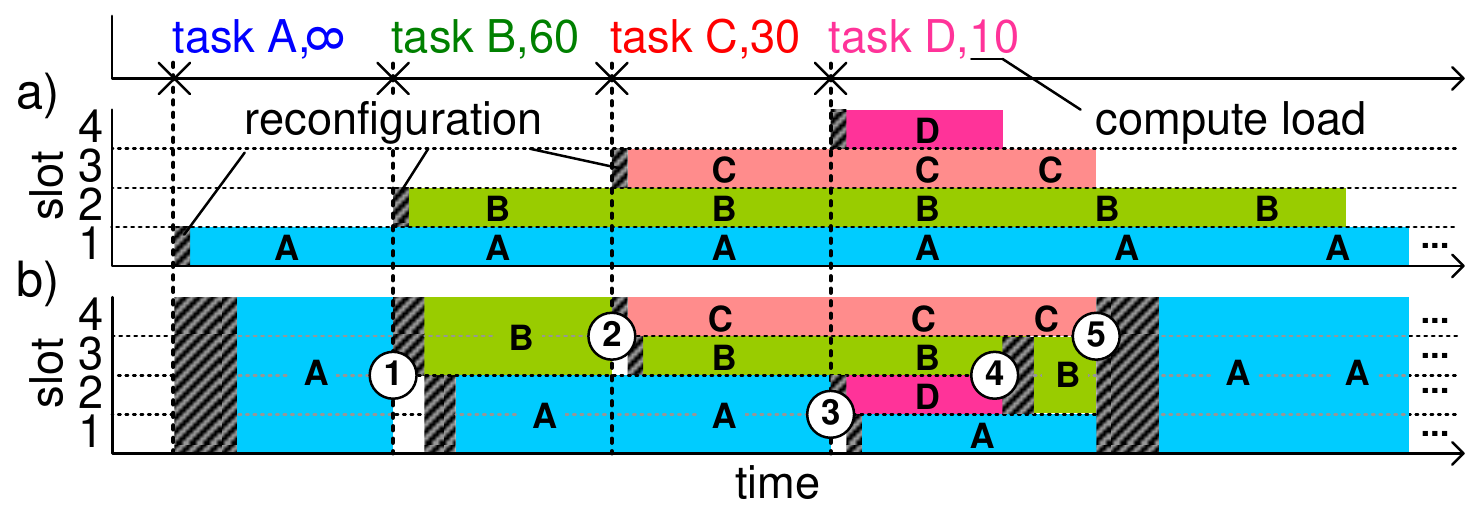}
    \caption{Resource allocation for kernels (tasks) A, B, C and D in time when using a) Standard fixed module scheduling and b) Resource-elastic scheduling on a 4 PR region FOS FPGA shell. The circled events highlight cases where resources are needed to accommodate new arriving tasks (\textcircled{1}, \textcircled{2}, \textcircled{3}) or cases where tasks complete (\textcircled{4}, \textcircled{5}).}
    \label{fig:resource_elasticity}
\end{figure}

Consequently, a single task execution call may execute on multiple accelerators in parallel or use an implementation alternative or share an accelerator with other tasks in time, during its execution lifetime. All these modes can be arbitrarily used without an application being aware of other tasks and types of accelerators it executes on. Hence, the runtime system is responsible for dynamically arranging the loading and unloading of these heterogeneous accelerators (written in C, C++, OpenCL or RTL) transparently from the user. 
Figure~\ref{fig:resource_elasticity} shows an example of how such resource allocation can allow maximising the FPGA utilisation and performance compared to standard fixed-module scheduling policies. 



\section{Evaluation}
\label{sec:evaluation}

    There are five primary dimensions onto which we can evaluate an FPGA operating system: 1)~FPGA resource overhead, 2)~software stack overhead, 3)~available memory performance, 4)~level of modularity and 5)~application performance. We detail and discuss the performance of FOS on each dimension for Ultra-96 and ZCU102 boards.

    \subsection{FPGA Resource Overhead}
        \subsubsection{FPGA shell}
        The resources used by an FPGA shell have a direct impact on the FPGA resources available for user hardware accelerators. Hence, it is important to minimise the overhead as much as possible. Table~\ref{tb-resources} shows the resources available for hardware acceleration when using FOS on ZCU102, Ultra-96, and UltraZed boards. 
        For the ZCU102 board (an MPSoC development kit from Xilinx) around 50\% of the resources are available for user acceleration whereas on a small IoT category Ultra-96 board it is about 80\%. This is because the layout of the ZUC102 chip is irregular, limiting the available resources when supporting relocatable modules (see Figure~\ref{fig:zcu102_static}).
        With a regular resource layout like in Ultra-96 (see Figure~\ref{fig-fos-impl}), we can maximise the resource allocation for relocatable modules and considerably reduce the resource overhead of a FOS shell. However, it is important to note that we do not entirely waste unused resources; they are available for future extensions as well as to implement other host system components, for example static accelerators or I/O functionalities.
        
        \begin{table}
        \centering
        \caption{Available resources for acceleration on the ZCU102 platform and the UltraZed \& Ultra96 platforms. The version on ZCU102 has 4 PR regions in total, while the other platforms provide 3 PR regions in total.}
        \small
        \begin{tabularx}{\linewidth}{X X X X}
            \toprule
            Resources on ZCU102  & Number of resources for 1 PR region & Chip utilisation per PR region (\%) & Total chip utilisation for accelerators (\%)\\
            \midrule
            CLB LUTs & 32640 & 11.70 & 46.80\\\midrule
            CLB Regs. & 65280 & 11.90 & 47.60\\\midrule
            BRAMs & 108 & 12.10 & 48.40\\\midrule
            DSPs & 336 & 13.30 & 53.20\\
            \bottomrule
            \toprule
            Resources on Ultra96 \& UltraZed  & Number of resources for 1 PR region & Chip utilisation per PR region (\%)& Total chip utilisation for accelerators (\%)\\
            \midrule
            CLB LUTs & 17760 & 25.17 & 75.51\\\midrule
            CLB Regs. & 35520 & 25.17 & 75.51\\\midrule
            BRAMs & 60 & 27.78 & 83.33\\\midrule
            DSPs & 96 & 26.67 & 80\\
            \bottomrule
        \end{tabularx}
        \label{tb-resources}
        \end{table}
        
        \subsubsection{Bus Virtualisation}
        
        In order to achieve modularity at the hardware interface layer, bus virtualisation is vital. However, to be able to dynamically load the interconnect wrappers implies pre-allocation of a partial region. Table~\ref{tb-logical-virt} shows the overhead of this pre-allocation of resources.
        We can identify that the unused resources are only about 448LUTs (18\% of pre-allocation) when we change the interconnection interface and protocol considerable, such as from AXI Stream to AXI master and slave.
        In particular, for large FPGAs this overhead is negligible. However, when using small FPGAs such as Ultra-96 or performing minor changes to the interface (e.g., changing bus width), the overhead of dynamic bus virtualisation is considerable.
        Hence, in such scenarios, we recommend using compile-time bus wrappers.
        
        \begin{table*}
        \centering
        \caption{Resource overheads for bus virtualisation at the logical and physical levels.}
        \small
        \resizebox{\textwidth}{!}{%
        \begin{tabularx}{\textwidth}{XXXlll}
            \toprule
            \multirow{2}{3cm}{Module \\Interface} & \multirow{2}{3cm}{Shell Interface} & \multirow{2}{3cm}{Bus adaptor's services} & \multicolumn{3}{c}{Resource overhead}\\
             &  & & Primitives & Logical Level & Physical Level\\
            \midrule
            \multirow{3}{3cm}{32-bit AXI-Lite \\\& 32-bit AXI4\\ Master} & \multirow{3}{3cm}{32-bit AXI-Lite \\\& 128-bit AXI4\\ Master} & \multirow{3}{3cm}{AXI \\Interconnect} & LUTs & 153 & 2400\\\cmidrule(lr){4-6}
            & &  & FFs & 284 & 4800\\\cmidrule(lr){4-6}
            & & & BRAMs & 0 & 12\\
            \midrule
            \multirow{3}{3cm}{32-bit AXI -Lite \\\& 32-bit AXI\\ Stream} & \multirow{3}{3cm}{32-bit AXI-Lite \\\& 128-bit AXI4\\ Master} & \multirow{3}{3cm}{Control reg., \\AXI MM2S, \\\& AXI DMA} & LUTs & 1952 & 2400\\\cmidrule(lr){4-6}
            & & & FFs & 2694 & 4800\\\cmidrule(lr){4-6}
            & &  & BRAMs & 2.5 & 12\\
            \bottomrule
        \end{tabularx}
        }
        \label{tb-logical-virt}
        \end{table*}
    
    \subsection{Software Stack Overhead}
    
    The software stack of an FPGA OS incurs two types of latencies: compile-time and runtime. Compile-time latency relates to the time taken to compile hardware accelerators with all the additional constraints for partial reconfiguration in the implementation phase, while the other relates to the overhead caused by intermediate layers in the software stack during accelerator execution.
    
    \subsubsection{Compilation Latency}
        
\begin{figure}
        \hfill
          \subfloat[Xilinx PR compilation flow result.]{
        	\begin{minipage}{0.48\textwidth}
        	    \centering
                \includegraphics[width=0.7\linewidth]{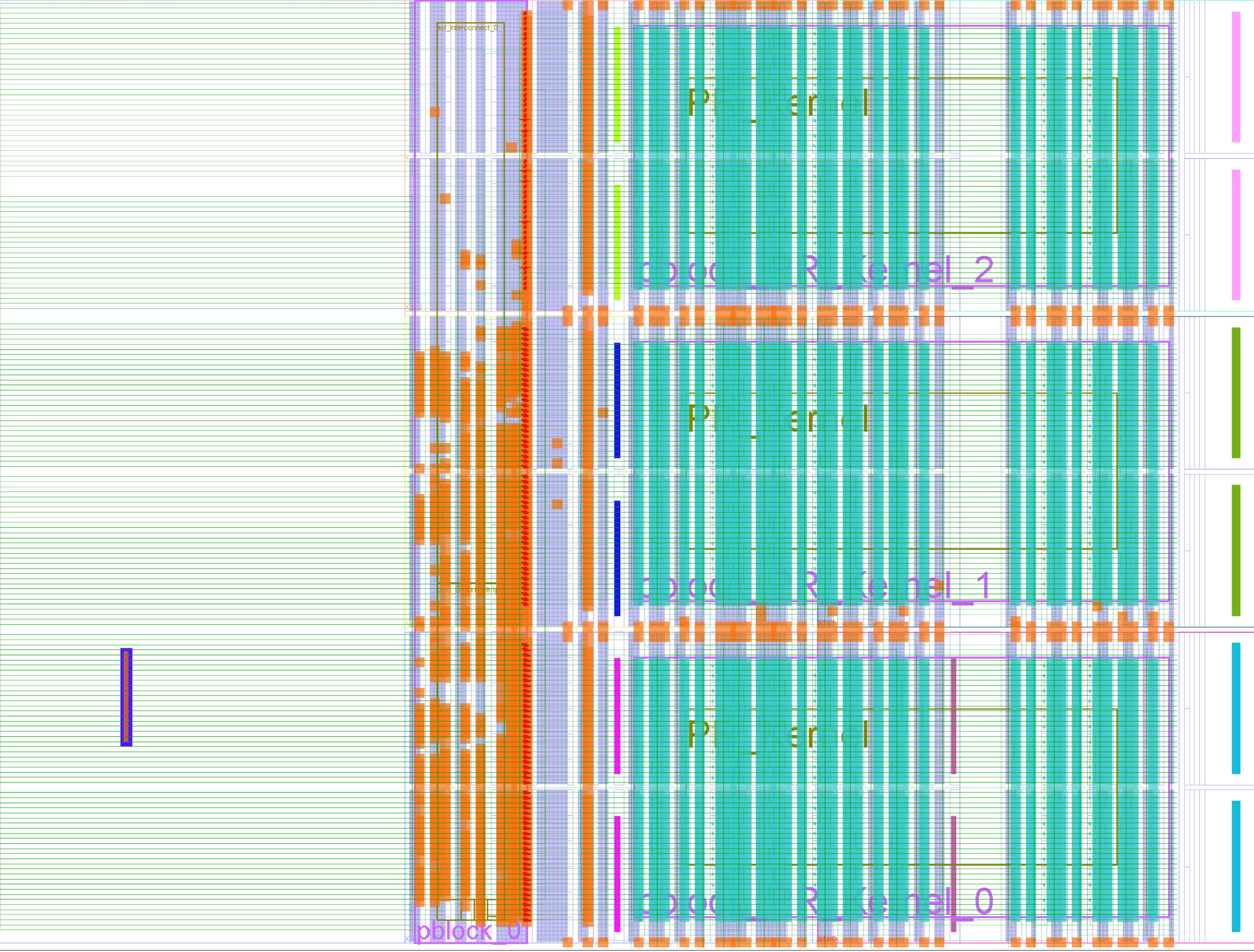}
        	\end{minipage}}
        \hfill
          \subfloat[FOS compilation flow result.]{
        	\begin{minipage}{0.49\textwidth}
        	    \centering
                \includegraphics[width=0.7\linewidth]{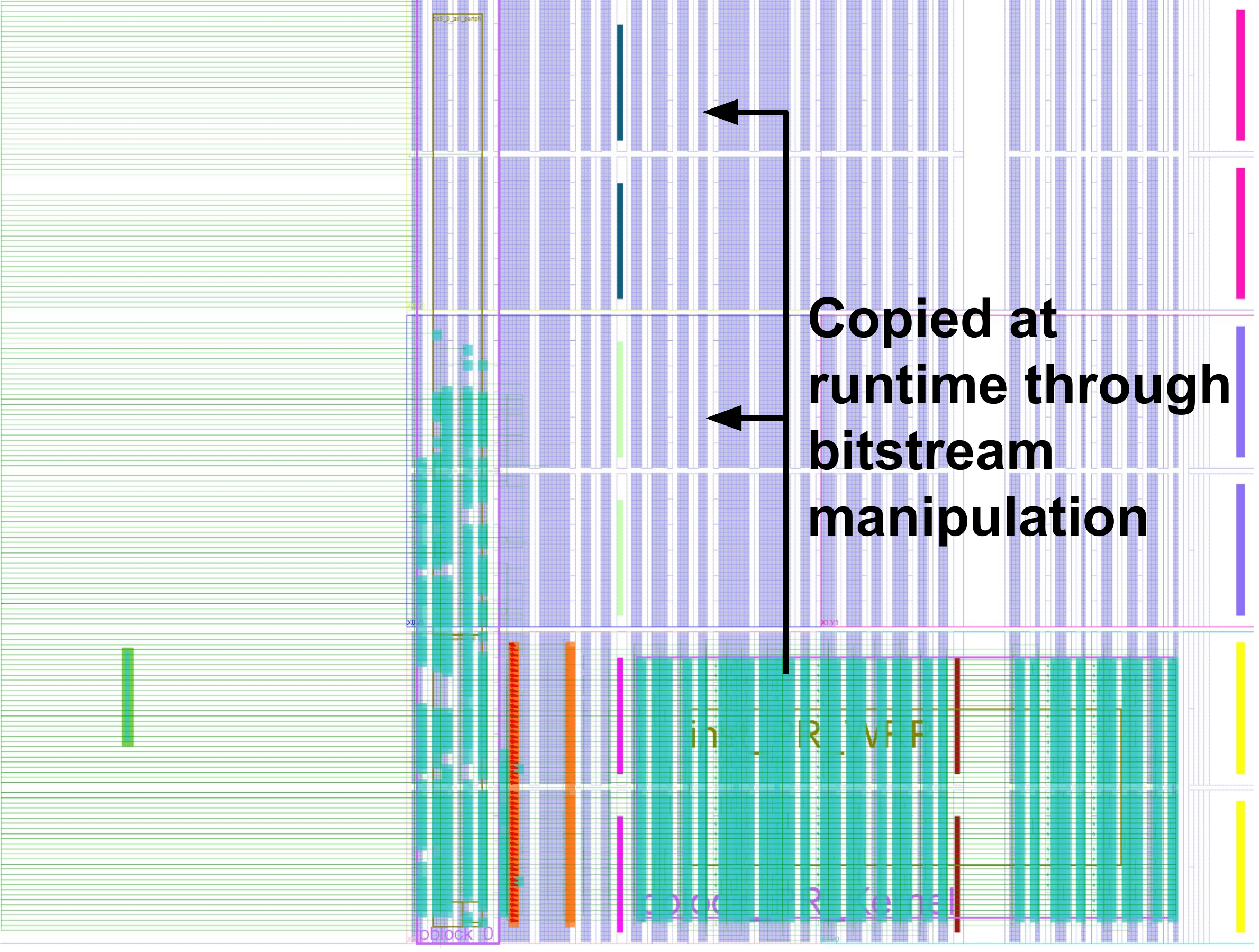}
        	\end{minipage}}
        \hfill
        \caption{Place and route result of Black Scholes accelerator~\cite{black_scholes} for Xilinx PR flow and FOS. Note, FOS generates relocatable bitstream with the help of BitMan~\cite{bitman} for other PR regions at runtime.}
        \label{fig:compilation_design_black_scholes}
        \end{figure}

        The standard Xilinx partial reconfiguration (PR) flow requires compiling both static and accelerator designs together and, more importantly, it needs to perform place and route (P\&R) and generate a new bitstream for each partial region.
        This leads to additional latency compared to our decoupled compilation flow, where we first generate a \textit{full-static} bitstream with Vivado and then a \textit{relocatable} partial bitstream with BitMan~\cite{bitman} for all regions. 
        To measure this, we used accelerators of three different types of size: sparse (AES), medium (Normal est.~\cite{spector}) and dense (Black Scholes~\cite{black_scholes}). The utilisation for each is 33\%, 63\% and 81\%, respectively. Figure~\ref{fig:u96_modules} shows the AES and Normal est. modules while the Black Scholes module is shown in Figure~\ref{fig:compilation_design_black_scholes} with a comparison to design generated by Xilinx PR flow. 
        
        Table~\ref{tab:compilation_lat} shows the latency breakdown for place and route and bitstream generation for both Xilinx PR flow and FOS compilation flow on Ultra-96. 
        The results show that 
        per region P\&R latency is higher for FOS as it adds additional constraints for relocatability, however, when compiling for multiple regions (i.e. 3 for Ultra-96 platform) it outperforms traditional compilation flow by up to 2.34$\times$ if a single slot accelerator can be duplicated to scale up the system. 
        Overall, when increasing the number of partial regions, the compilation flow latency of FOS stays constant, whereas the latency of the Xilinx PR flow increases linearly.
        This relationship turns into an exponential increase in compilation time when compiling several applications with standard Xilinx PR flow, as it needs to compile \textit{each module for each partial region}. Hence, it is important to adopt a scalable and modular PR flow when targeting multiple applications and shell versions which are commonly used in cloud environments.
        
        \begin{table}
        \centering
        \caption{Total place and route (P\&R), and bitstream generation latency for AES, Normal Est.~\cite{spector}, and  Black Scholes accelerator~\cite{black_scholes} when compiling for all three partial regions on Ultra-96 shell. Evaluation is conducted using Vivado 2018.2.1 on Intel core i7-4930K CPU running at 3.4 GHz with 64 GB of RAM.}
        \label{tab:compilation_lat}
        \small
        \resizebox{\textwidth}{!}{%
        \begin{tabular}{lllllllll}
        \toprule
        \multirow{2}{*}{Applications} & \multirow{2}{*}{\begin{tabular}[c]{@{}l@{}}Region\\Util.\end{tabular}} & \multicolumn{3}{c}{Xilinx PR}      & \multicolumn{3}{c}{FOS}            & \multirow{2}{*}{Speed Up} \\ \cmidrule(lr){3-8}
                                      &                                                                              & P\&R (s) & Bitgen (s) & Total (s) & P\&R (s) & Bitgen (s) & Total (s) &                           \\ \midrule
        AES                           & 33\%                                                                         & 429.40    & 176.19     & 605.59    & 284.18    & 64.06      & 348.24    & 1.74 $\times$             \\ \midrule
        Normal Est.                   & 63\%                                                                         & 747.75    & 201.21     & 948.96    & 387.41    & 70.09      & 457.50    & 2.07 $\times$             \\ \midrule
        Black Scholes                 & 81\%                                                                         & 1296.26   & 231.27     & 1527.53   & 574.56    & 77.11      & 651.67    & 2.34 $\times$             \\ \bottomrule
        \end{tabular}
        }
        \end{table}

    \subsubsection{Runtime Execution Overhead}
    
    The runtime overhead occurs because of the four main steps performed when using the FOS software stack: i)~initialisation of gRPC server, ii)~parsing of JSON files for accelerators and shells, iii)~gRPC call to the daemon and iv)~scheduling latency.
    Table~\ref{tab:runtime_overhead} details the latency of each step. 
    The first two steps (i and ii) have heavy dependencies on I/O as they use the network and file system, respectively. This leads to latencies in the range of milliseconds. However, this overhead is amortised over time as we perform steps i) and ii) only once at system start. The gRPC call to the daemon goes through many levels of the Linux stack (processes) before reaching the FOS multi-tenancy daemon, leading to a latency of about a millisecond. We can speed up the gRPC call by reducing the Linux timer interrupt period to achieve a better response time from the Linux kernel for quick turn around between processes if required.
    The scheduling latency is in the range of micro-seconds and comparable to standard CPU schedulers. Moreover, the scheduler is event driven and executed only when an accelerator finishes or a new acceleration request arrives (due to its cooperative nature) rather than at every timer interrupt like preemptive scheduling. 
    
    \begin{table}
    \centering
    \caption{Execution overhead caused by various software layers.}
    \label{tab:runtime_overhead}
    \small
    \begin{tabular}{ll}
    \toprule
    Software Layer      & Latency (ms) \\ \midrule
    Initialize gRPC (once)     & 12.20        \\ \midrule
    JSON parsing (once)        & 2.27        \\ \midrule
    gRPC Call to Daemon & 0.71         \\ \midrule
    Scheduler           & 0.02         \\ 
    \bottomrule
    \end{tabular}
    \end{table}
    
    \subsection{Memory Performance}
    
        One important characteristic of an FPGA operating system on the hardware acceleration is the memory bandwidth it can provide given the shell implementation (interconnect to the memory). Hence, we evaluated the available throughput on different AXI ports made available to partial regions as well as their combined throughput on Ultra-96 and ZCU102 boards with varying burst sizes from the PL using the memory evaluation kit~\cite{manev_mem_test}.
        
\begin{figure}
            \centering
            \includegraphics[width=0.68\linewidth]{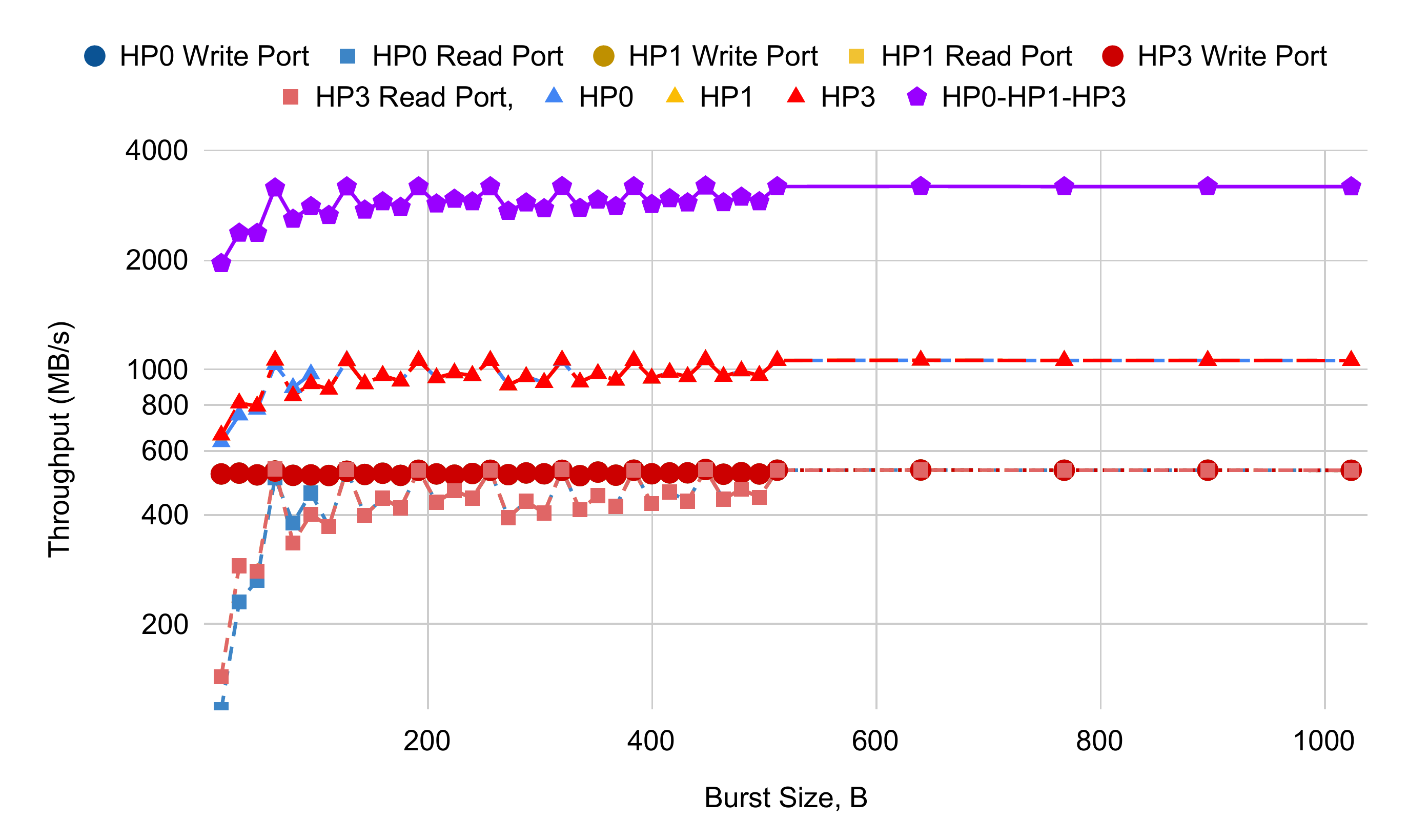}
            \caption{Memory throughput for varying burst sizes on the duplex AXI ports (HP0, HP1, and HP3) available to the hardware accelerators in PR regions 0 to 2 of Ultra-96 FOS platform at 100 MHz.}
            \label{fig:u96_mem}
        \end{figure}

        Figure~\ref{fig:u96_mem} shows the breakdown of the read and write throughput of each AXI port and the total accessible bandwidth for the Ultra-96 board, respectively. On average, there is an even split of read and write bandwidth with a throughput of 530~MB/s for each read and write operation. The aggregated read-write throughput of the individual AXIs is about 1060~MB/s and, when activated at the same time, all available AXI ports collectively achieve up to 3187~MB/s. This translates to about 25\% and 74\% of the theoretical DDR peak throughput when using AXI ports individually and concurrently, respectively.
    
\begin{figure}
            \centering
            \includegraphics[width=0.68\linewidth]{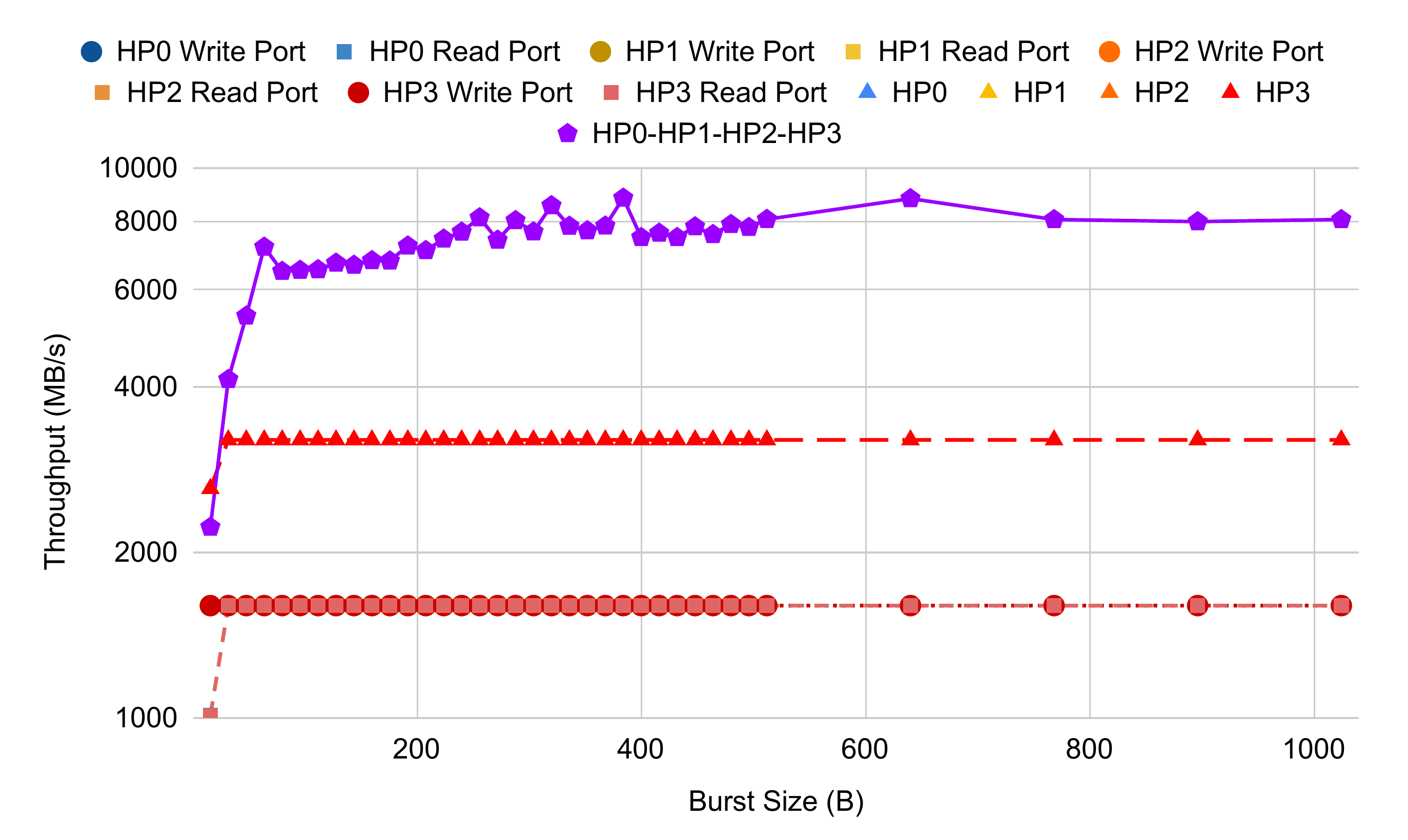}
            \caption{Memory throughput for varying burst sizes on the duplex AXI ports (HP0 to HP3) available to the hardware accelerators in PR regions 0 to 3 of ZCU102 FOS platform at 100 MHz.}
            \label{fig:zcu102_mem}
        \end{figure}

        Figure~\ref{fig:zcu102_mem} shows the breakdown of read and write throughput on the ZCU102 board. Each AXI port achieves a throughput distribution between read and write transactions 1600~MB/s each. 
        The total throughput is 3200~MB/s for individual AXI ports and 8804~MB/s when using all AXIs together, as shown in Figure~\ref{fig:zcu102_mem}.
        We expect that sub-linear improvement in the total throughput when using all AXI ports (HP0-3) concurrently is caused by the row pollution and AXI interconnect multiplexing in the memory controller.
        
    \subsection{Quantifying Modularity}
    
        Compared to the standard FPGA development flow, where a change in the shell can mean recompilation of all system components (both hardware and software), the modular FOS FPGA development flow provides the freedom to \textit{update individual components} without recompilation of other components which may take hours~\cite{petalinux, feist2012vivado, khawaja2018armophos}. 
        This means that FOS only has to pay compilation and re-initialisation latency (as shown in Table~\ref{tab:in-field_update}) for the component which is being changed given that it does not change the interfaces defined between them.
        
        From Table~\ref{tab:in-field_update}, we can identify that this avoidance of recompilation allows changing the shell at runtime with additional functionalities or bug fixes costs less than 21~ms on Ultra-96 (IoT device) and 99~ms on ZCU102 (Xilinx MPSoC development kit)\footnote{This assumes that the shell bitstream is pre-compiled as the recompilation of shell itself is unavoidable for all systems.}. Similarly, swapping an accelerator implementation is easy and includes only the partial reconfiguration latency, because FOS provides generic drivers. In contrast to FOS, the standard flow would require generating/writing new drivers and re-installing them separately. 
        Changing the kernel involves the biggest re-initialisation latency, as this requires a system reboot which takes 66 seconds in total, including I/O setup (for keyboard, mouse, Wi-Fi, monitor and webcam) on Ultra-96. However, this still avoids the need to update user software binaries for re-integration as required in the standard PetaLinux flow~\cite{petalinux}.
        
        Overall, modularity of FOS removes the recompilation latencies and re-development steps which cost in the range of hours~\cite{khawaja2018armophos} and bring down the component change latency by two-orders of magnitude compared to the standard development flow while supporting all the features required from an FPGA OS.
    
        \begin{table}
        \centering
        \caption{Re-initialisation latencies for component change on FOS platforms.}
        \label{tab:in-field_update} 
        \small
        \begin{tabular}{l l l}
        \toprule
        Component Updated & U-96 Latency (ms) & ZCU102 Latency (ms) \\ \midrule
        Accelerator      & 3.81                                   & 6.77                                    \\ \midrule
        Shell            & 20.74                                  & 98.4                                    \\ \midrule
        Runtime          & 15.2                                     & 15.2                                   \\ \midrule
        Kernel     & 66000                                  & 15760                                   \\ \bottomrule
        \end{tabular}
        \end{table}
    
    \subsection{Application Case Study}
    
        We evaluated a case-study in two different environments: 1)~single-tenant but multiple partial regions and 2)~multi-tenant with dynamic offloading. All the accelerators used in this case study operate at 100 MHz.
    
        \subsubsection{Single-tenant with Multiple Partial Regions}
        
\begin{figure}
            \centering
            \begin{minipage}{.49\textwidth}
            \includegraphics[width=\linewidth]{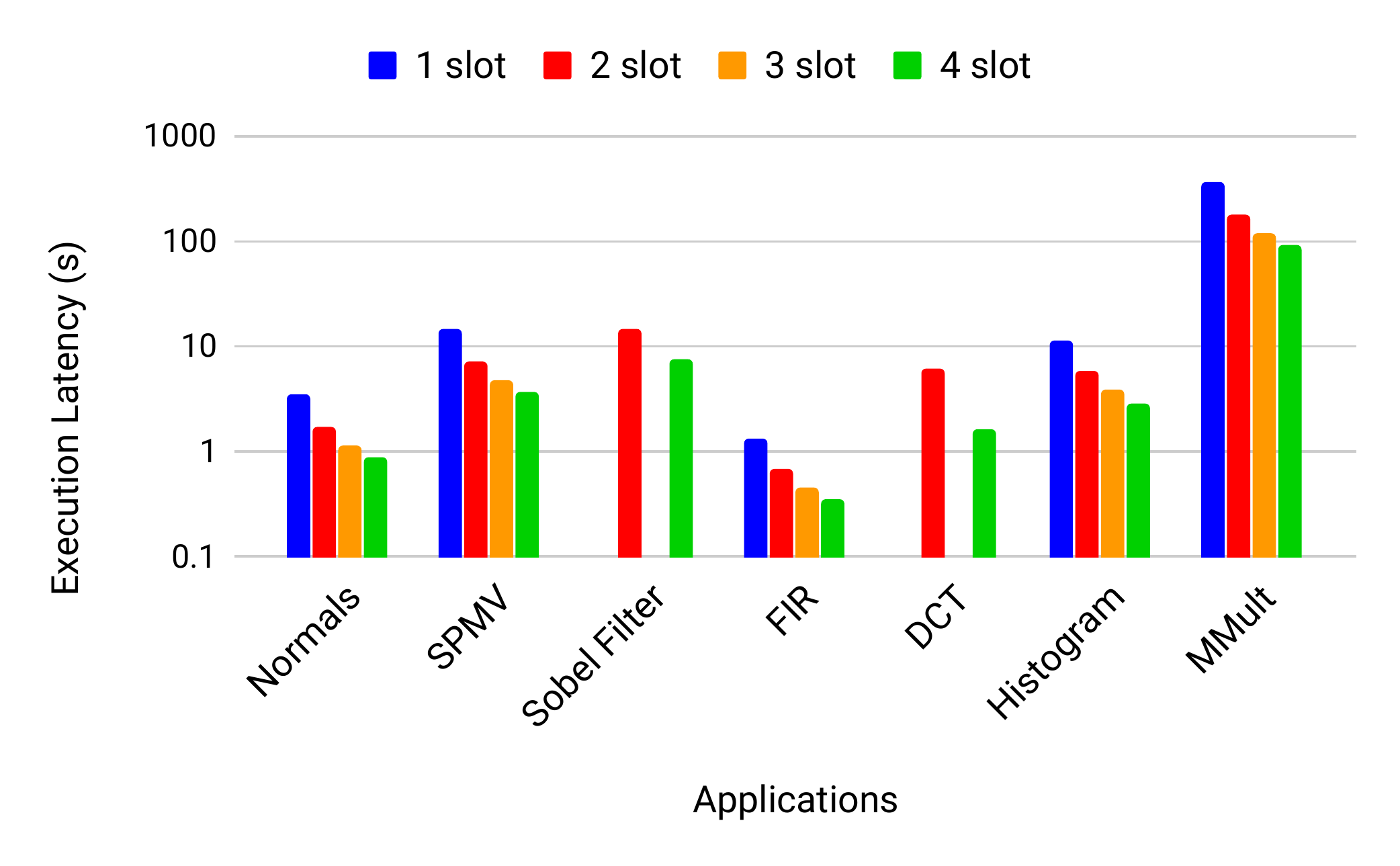}
            \caption{Execution latencies of accelerators from the Spector benchmark suite on ZCU102 platform.}
            \label{fig:zcu102_spector}
            \end{minipage}
            \hfill
            \begin{minipage}{.49\textwidth}
            \includegraphics[width=\linewidth]{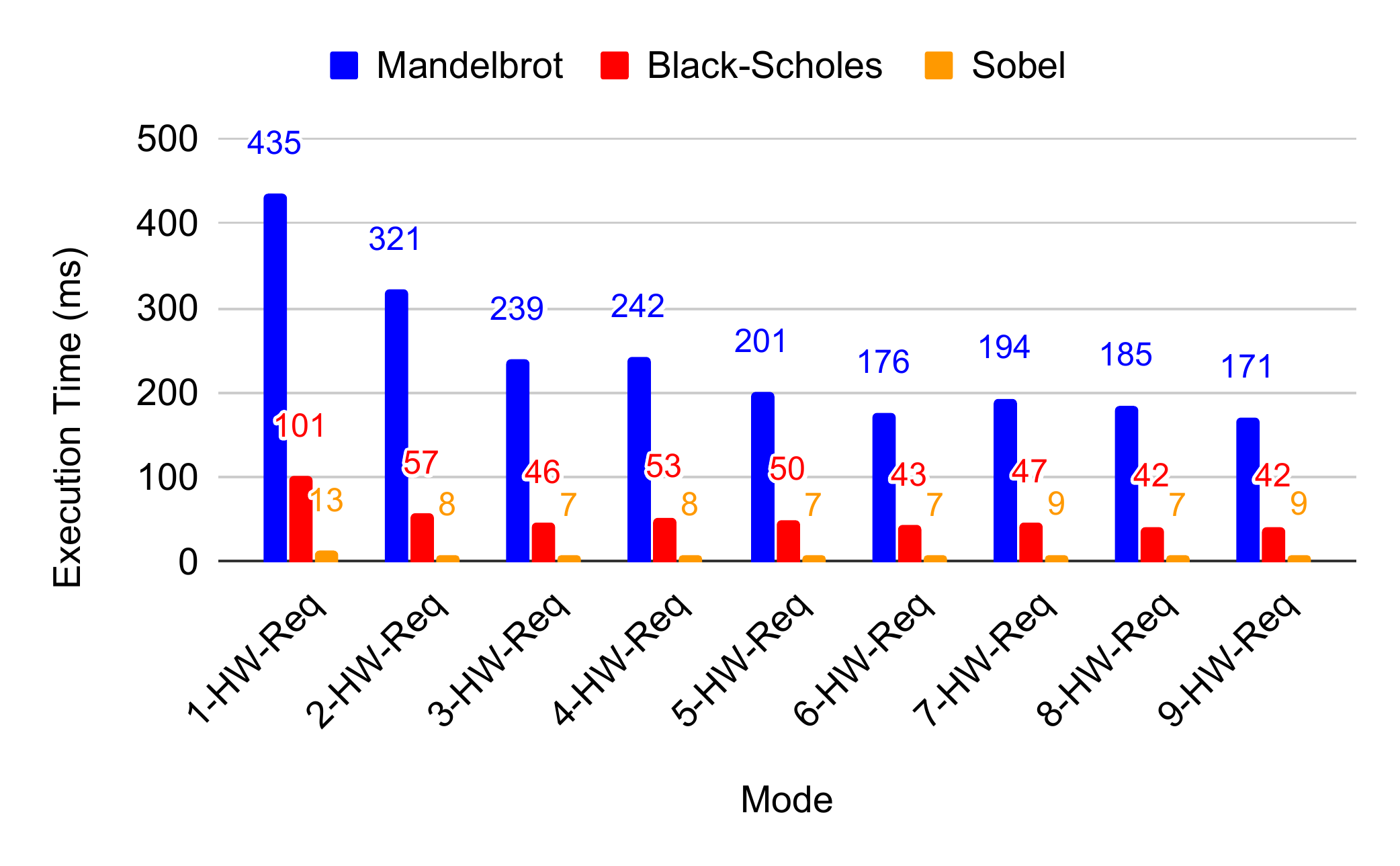}
            \caption{Execution latencies of Mandelbrot, Black Scholes (European option)~\cite{black_scholes}, and Sobel~\cite{xilinx_sdaccel_examples} when executing concurrently with varying amount of hardware requests on Ultra-96 platform.}
            \label{fig:standalone_apps}
            \end{minipage}
        \end{figure}

        To evaluate the benefits of multiple partial regions and the ability to replicate accelerators dynamically, we selected OpenCL accelerators from the Spector benchmark suite~\cite{spector} for two reasons: 1)~it comprises a wide range of application behaviour from multiple domains, and 2)~it is annotated with HLS pragmas to generate implementation alternatives without modifying the source code. Figure~\ref{fig:zcu102_spector} shows the results of the execution latencies with a varying amount of resources available for acceleration on the ZCU102 platform. Most of the benchmark applications show an almost \textit{linear} performance improvement when replicated across multiple partial regions. In particular, DCT benefits from the ability to switch to bigger module implementation (by using more data buffers and a larger unrolling factor) and achieves a \textit{super-linear} performance improvement of 3.55$\times$ for 2$\times$ resources. 
        
\begin{figure}
            \centering
            \begin{minipage}{.49\textwidth}
            \includegraphics[width=\linewidth]{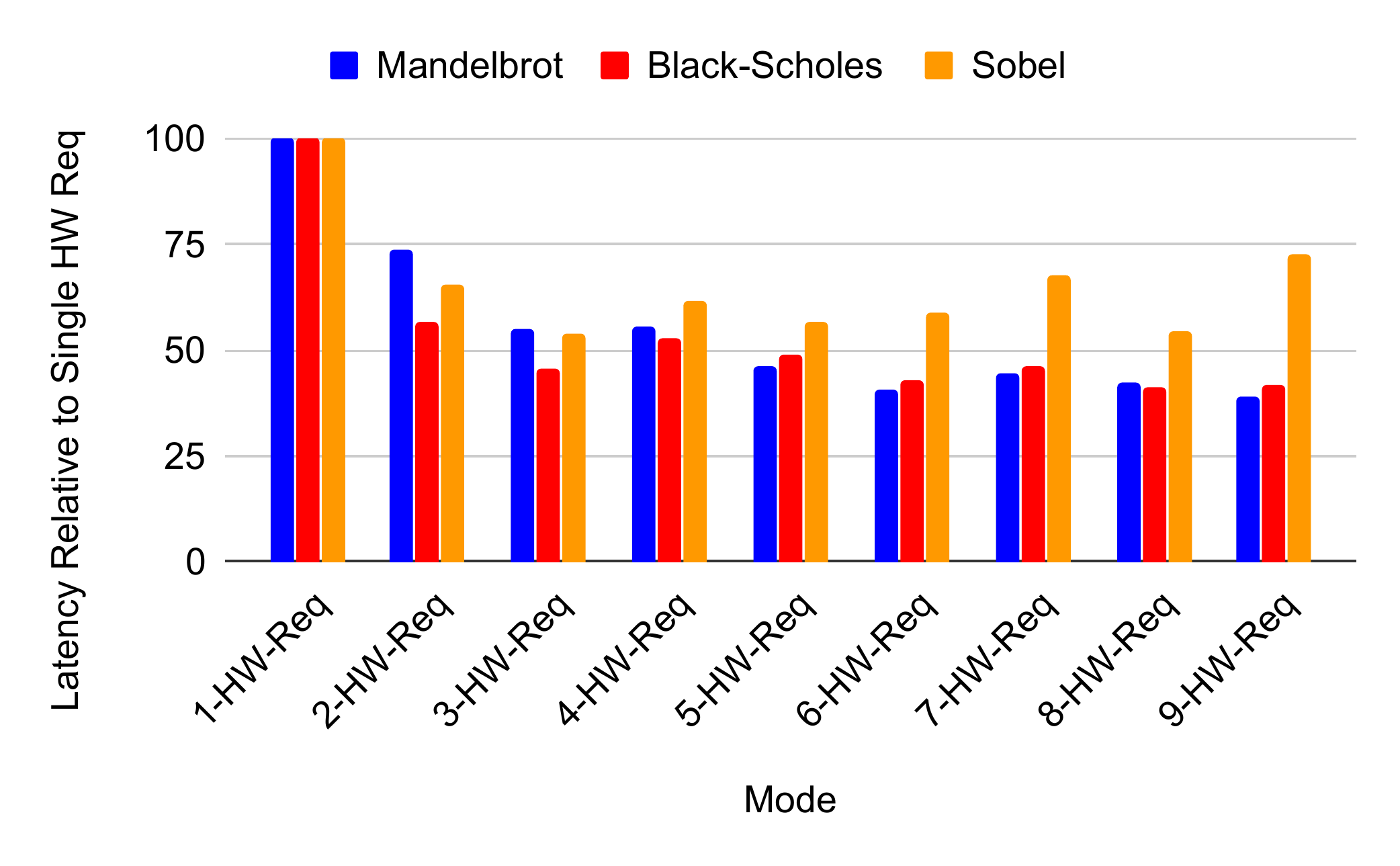}
            \caption{Relative execution latencies of Mandelbrot, Black Scholes (European option)~\cite{black_scholes}, and Sobel~\cite{xilinx_sdaccel_examples} applications when exposing varying amount of parallelism to process a frame on Ultra-96 platform.}
            \label{fig:standalone_apps_relative}
            \end{minipage}
            \hfill
            \begin{minipage}{.49\textwidth}
            \includegraphics[width=\linewidth]{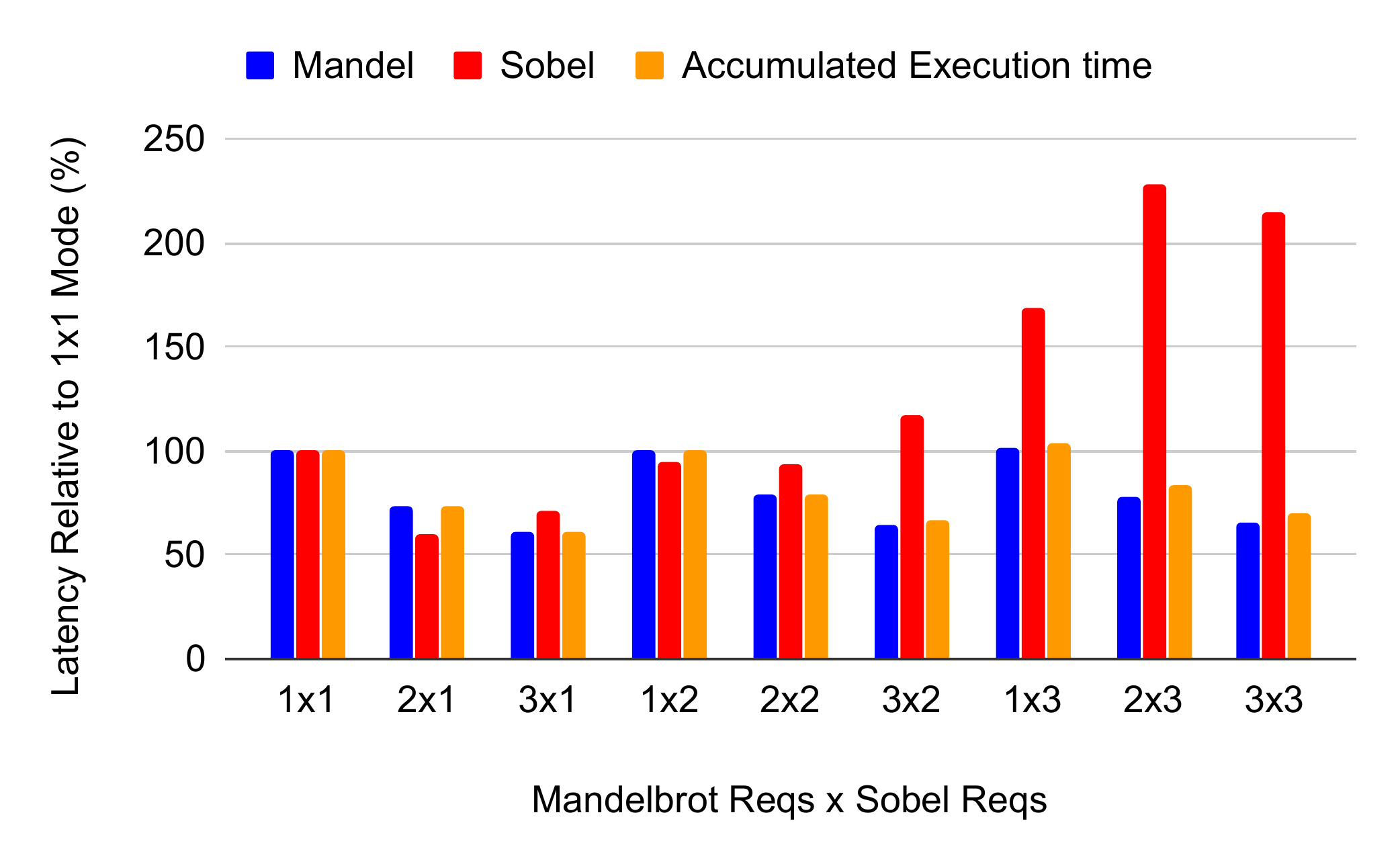}
            \caption{Relative execution latencies of Mandelbrot and Sobel~\cite{xilinx_sdaccel_examples} when executing concurrently with varying amount of HW requests on Ultra-96 platform}
            \label{fig:MandelxSobel_relative}
            \end{minipage}
        \end{figure}

        To understand the effects of exposing more parallelism than available on the FPGA platform, we conducted the experiments on Ultra-96 using compute-bound (Mandelbrot and Black-Scholas~\cite{black_scholes}) and memory-bound (Sobel~\cite{xilinx_sdaccel_examples}) applications which are more sensitive to reconfiguration overhead due to their smaller execution latencies than Spector benchmarks~\cite{spector}. Figure~\ref{fig:standalone_apps} shows the results in which the number of requests dictates the amount of parallelism exposed to the runtime. The performance improves almost linearly until it hits the number of available partial regions (3 in the case of Ultra-96 platform), after which the performance tends to stagnate (see Figure~\ref{fig:standalone_apps_relative}). This is because the scheduler uses time multiplexing to provide the illusion of an unlimited number of regions, leading to behaviour similar to multi-threading on CPUs. In particular, cases, where the number of requests is a multiple of the number of physical accelerators, perform better than others as it avoids bottlenecks caused by the (leftover) pending requests at the end of execution.
        
        Overall, this highlights that FOS can help to improve performance even when using a single application along with its other benefits of modularity and developer productivity.
        
        
        \subsubsection{Multi-tenant Dynamic Offload}
        
        To understand the performance changes when using multiple applications at the same time, we execute the Mandelbrot and Sobel~\cite{xilinx_sdaccel_examples} applications concurrently on the Ultra-96 platform. Note that individual applications are not aware of other applications executing on the platform. In particular, the accelerators are written in C (mandelbrot) and OpenCL (sobel), demonstrate support for multiple different languages used concurrently.
        
        Figure~\ref{fig:MandelxSobel_relative} shows the execution latencies relative to 1-Mandel$\times$1-Sobel scenario with a varying amount of acceleration requests.
        As we can see, the execution latencies tend to decrease with an increase in the number of requests as in the standalone case. However, here the optimal performance is achieved at 3-Mandel$\times$1-Sobel rather than 3-Mandel$\times$3-Sobel. This is because of two reasons: 1)~adding more Sobel units reduces memory performance due to row-bank pollution and 2)~multiple request from different application induces rapid reconfiguration latencies. Regardless of this behaviour, it is important to note that if each application takes a greedy decision to request the highest amount of parallelism suited for the application based on the standalone results, the system can still achieve a near-optimal performance resulting in 46\% improvement over 1-Mandel$\times$1-Sobel by dynamically reallocating resources using the same accelerators. 

\section{Conclusion}


In this paper, we described the underlying concepts and abstraction layers involved in building a modular FPGA operating system that can adapt to dynamic workloads. The resulting FPGA operating system -- FOS, provides support for traditional as well as multi-tenancy environments with low overhead and easy to use software interfaces. The dynamic resource allocation capabilities of FOS, allows FOS to share multiple FPGA accelerators transparently between multiple users in the time and the spatial domain as well as the ability to switch between accelerator implementations on the fly.
Our evaluation shows that FOS can speed up the compilation time by up to 2.34x and avoid standard recompilation requirements to reduce the update latencies by over 100x with its modular FPGA development flow. The overheads caused by the hardware and software layers are minor and recovered by the ability to schedule resources dynamically in both single and multi-tenant environments. 
Overall, FOS directly caters the needs of upcoming FPGA systems which are being deployed at scale (cloud and edge) and allows systems to be more maintainable, adaptable and accessible, and benefiting both FPGA and application domain experts.
As a distinct feature, FOS provides an application-centric view to the developers, by hiding most of the complexity encountered when using a heterogeneous CPU-FPGA acceleration system with Linux back-end. 
With this, FOS is empowering a larger FPGA user community to implement complex and scalable heterogeneous systems.

\newpage
%
\bibliographystyle{unsrt}
\bibliography{main.bib}

\end{document}